\documentclass[12pt,a4paper]{article}

\usepackage[british]{babel}

\usepackage[a4paper,top=2cm,bottom=2cm,left=2.5cm,right=2.5cm,marginparwidth=1.75cm]{geometry}

\usepackage[style=apa, backend=biber]{biblatex} 
\DeclareLanguageMapping{british}{british-apa} 
\DeclareFieldFormat[article]{volume}{\apanum{#1}} 

\usepackage{amsmath}
\usepackage{graphicx}
\usepackage[colorlinks=true, allcolors=blue]{hyperref}
\usepackage{hyperref}
\usepackage[title]{appendix}
\usepackage{mathrsfs}
\usepackage{amsfonts}
\usepackage{booktabs} 
\usepackage{caption}  
\usepackage{threeparttable} 
\usepackage{algorithm}
\usepackage{algorithmicx}
\usepackage{algpseudocode}
\usepackage{listings}
\usepackage{enumitem}
\usepackage{chngcntr}
\usepackage{booktabs}
\usepackage{lipsum}
\usepackage{subcaption}
\usepackage{authblk}
\usepackage[T1]{fontenc}    
\usepackage{csquotes}       
\usepackage{diagbox}
\usepackage{comment}

\usepackage{subcaption}
\usepackage[justification=raggedright, singlelinecheck=false]{caption}
\usepackage{caption}
\usepackage{helvet}  
\usepackage{lmodern}  

\usepackage{setspace}
\usepackage{titlesec}
\titleformat{\section} 
  {\normalfont\Large\bfseries}{\thesection.}{1em}{}
  
\usepackage{lineno} 

\rightlinenumbers 

\linenumbers 

\newcommand{\subsubsubsection}[1]{%
  \vspace{\baselineskip}
  \noindent\textbf{#1\\}\quad
}
\usepackage{float}   
\usepackage{caption} 


\title{Adaptive Uncertainty-Guided Surrogates for Efficient phase field Modeling of Dendritic Solidification}

\author[1,2,*]{Eider Garate-Perez}
\author[1]{Kerman López de Calle-Etxabe}
\author[1]{Oihana Garcia}
\author[3]{Borja Calvo}
\author[1]{Meritxell Gómez-Omella}
\author[4]{Jon Lambarri}
\affil[1]{\small Intelligent Information Systems, Tekniker, Iñaki Goenaga 5, Eibar (20600) Spain}
\affil[2]{\small Faculty of Informatics UPV-EHU, Paseo Manuel de Lardizabal 1, Donostia (20700) Spain}
\affil[3]{\small Department of Computer Sciences and Artificial Intelligence, UPV-EHU, Donostia (20700) Spain}
\affil[4]{\small Advanced Manufacturing Technologies, Tekniker, Iñaki Goenaga 5, Eibar (20600) Spain}
\affil[*]{\small Corresponding author. \texttt{eider.garate@tekniker.es}}

\date{}  

\begin{document}
\maketitle

\begin{abstract}
The high computational cost of phase field simulations remains a major limitation for predicting dendritic solidification in metals, particularly in additive manufacturing, where microstructural control is critical. This work presents a surrogate model for dendritic solidification that employs uncertainty-driven adaptive sampling with XGBoost and CNNs, including a self-supervised strategy, to efficiently approximate the spatio-temporal evolution while reducing costly phase field simulations. The proposed adaptive strategy leverages model uncertainty, approximated via Monte Carlo dropout for CNNs and bagging for XGBoost, to identify high-uncertainty regions where new samples are generated locally within hyperspheres, progressively refining the spatio-temporal design space and achieving accurate predictions with significantly fewer phase field simulations than an Optimal Latin Hypercube Sampling optimized via discrete Particle Swarm Optimization (OLHS-PSO). The framework systematically investigates how temporal instance selection, adaptive sampling, and the choice between domain-informed and data-driven surrogates affect spatio-temporal model performance. Evaluation considers not only computational cost but also the number of expensive phase field simulations, surrogate accuracy, and associated $CO_2$ emissions, providing a comprehensive assessment of model performance as well as their related environmental impact.
\end{abstract}


\section{Introduction}\label{sec:intro}


The ecological awareness that has emerged in recent decades has reshaped manufacturing, prioritizing durability, sustainable materials, and production processes that reduce energy consumption and waste, while emphasizing recyclability and reusability (\cite{jarfors_functionally_2024, panagiotopoulou_critical_2022}). To this transformation is added the other design conflicts that until now were found in manufacturing: improving the physical properties of the materials and maximizing the economic performance of the products (\cite{fassi_towards_2023}). This convergence highlights the importance of designing products that simultaneously meet ecological, functional, and economic requirements.


In this context, computational models offer a significant advantage by capturing the underlying physical relationships of the phenomena involved (\cite{wang_hybrid_2022}), avoiding unnecessary material waste and reducing the associated economic costs (\cite{nyamekye_sustainability_2024}). This approach not only accelerates innovation but also aligns with sustainable practices.

However, traditional and accurate physical models capable of simulating the properties associated with a design are often computationally expensive, particularly when searching for optimal configurations across large design spaces (\cite{duquesnoy_machine_2023, wurth_physics-informed_2023, xue_reliability-based_2024}). Artificial Intelligence (AI) and Machine Learning (ML) have emerged as powerful alternatives to complex simulations, offering the ability to rapidly and reliably predict material properties (\cite{wang_hybrid_2022, usman_prediction_2024}). This more efficient prediction enables the optimization of design problems that would otherwise be prohibitive when relying solely on conventional computational models. 

This approach, commonly termed a surrogate model (\cite{azarhoosh_review_2025, ravutla_effects_2025, kaminsky_efficient_2021}), aims to replace complex computational models or experiments with ML approximations that minimize accuracy loss while enabling optimization, design exploration, and performance improvement at manageable computational cost (\cite{khatouri_metamodeling_2022}). Surrogate models are generally trained on reduced datasets that capture the essential characteristics of the phenomenon, generated either by the original model or derived from experimental data (\cite{forrester_recent_2009}). The construction of such datasets, known as sampling (\cite{di_fiore_active_2024}), is critical to ensuring representativeness while avoiding excessive computational expense.

In the literature, two main sampling strategies are distinguished: classical sampling, which is performed at the initial stage when limited information about the phenomenon is available, and therefore focuses on ensuring adequate exploration of the design space; and adaptive sampling, which iteratively integrates with surrogate model training to exploit model information and thereby enhance both exploration and exploitation of the space (\cite{liu_survey_2018, cvetkovic_surrogate_2017, samadian_application_2025}). Many studies have shown that adaptive sampling is able to achieve more accurate surrogate models with less data than classical sampling, and therefore, the training of these is more efficient with particularly expensive models or experiments (\cite{viana_surrogate_2021}).

Building on these advances, it becomes possible to address domain-specific challenges that traditionally rely on expensive simulations. One such challenge is the prediction of dendritic solidification geometries in metals, a complex phenomenon whose accurate modeling is computationally demanding (\cite{bhagat_modeling_2023}). These microstructural patterns are directly linked to the physical and mechanical properties of the solidified material (\cite{kurz_progress_2021}), and their accurate characterization is essential for ensuring the performance of the final component. The importance of this problem is intensified in additive manufacturing (AM), where metals undergo rapid solidification processes that critically determine the reliability of the produced parts (\cite{yu_impact_2022}). AM technologies, recognized for their precision, reduced material waste, and environmental advantages (\cite{javaid_role_2021, jiang_machine_2022}), require models capable of capturing solidification phenomena to align with both ecological and functional demands. 

Among the computational approaches employed to address this challenge, the phase field model stands out as one of the most widely used (\cite{tang_phase_2022, dobravec_coupled_2023, zeng_phase-field_2024}). This method enables the simulation of dendritic solidification geometries in 1D, 2D, or 3D, providing a detailed description of microstructural evolution during solidification (\cite{mao_anisotropic_2024, seiz_simulation_2023, ji_phase-field_2025}). Although higher-dimensional phase field models yield greater accuracy in simulating dendritic solidification, they also entail a steep increase in computational cost. In additive manufacturing, this makes surrogate models particularly valuable, as they approximate phase field behavior with far lower computational demand (\cite{lee_recent_2023, choi_accelerating_2024, viardin_automatic_2025}). In this context, adaptive sampling is crucial for efficient surrogate modeling. By guiding data acquisition based on model predictions, it reduces the computational cost of high-fidelity simulations while ensuring that key features of dendritic solidification are accurately captured. Nevertheless, these approaches still rely on an initial dataset generated by the phase field model. Together, this strategy enables the construction of precise surrogate models with fewer simulations, balancing accuracy and efficiency in the exploration of solidification processes.

One of the biggest challenges when replacing phase field models is dealing with their inherent spatio-temporal nature (\cite{affenzeller_adaptive_2024, migdady_adaptive_2025}). Most of the literature focuses on using several temporal instances to predict the final stage of solidification, as well as other properties related to the evolution of the microstructure. For that reason, the challenge lies in accurately capturing both the temporal progression and the spatial complexity of the system while minimizing reliance on the original phase field model during training and prediction.


In the literature, different strategies have been proposed to address the spatio-temporal complexity of phase field simulations. For instance, \cite{montes_de_oca_zapiain_accelerating_2021} employs the first 80 temporal instances to predict the last 5, applying a two-point statistics transformation followed by PCA and an LSTM to estimate the final microstructure. \cite{peivaste_machine-learning-based_2022} predicts subsequent instances using an encoder–decoder (U-net) architecture, while \cite{yan_machine_2024} combines PCA, autoencoders, and LSTM networks to infer the final state from 50 instances. Similarly, \cite{ahmad_accelerating_2023} investigates the number of temporal instances required to predict the final microstructure using a convolutional neural network coupled with LSTM layers. Although these approaches successfully reduce the temporal dimensionality of the problem, they still rely on phase field simulations to generate the predictor instances. Consequently, the choice and number of temporal snapshots become critical, as they directly determine the computational cost and efficiency of the overall framework. This highlights the need for strategies that not only learn from reduced temporal information but also minimize dependence on high-fidelity simulations when constructing predictive models.

In this framework, surrogate models play a central role in capturing the spatio-temporal correlations inherent to phase field simulations while alleviating their computational burden. Consequently, the selection of predictive instances becomes a critical design decision, as it directly affects model accuracy, complexity, and convergence, while also determining the degree to which dependence on the high-fidelity simulator can be reduced. When instances must be generated by the real physical model up to a time $t_f$, computational savings are restricted to the interval between $t_f$ and the final solidification stage.

Beyond these key aspects, it is also essential to evaluate whether surrogate models truly deliver efficiency in a broader sense. When proposing such models to design more effective manufacturing processes—whether to enhance functionality, reduce economic costs, or mitigate environmental impact—it becomes crucial to measure and compare their performance against sustainability constraints.

The present work introduces an adaptive sampling methodology to train a surrogate model for dendritic solidification in metals, with a focus on additive manufacturing. Three main objectives guide this study: (i) to analyze the effect of temporal training instance selection on model performance, training cost, and potential computational savings relative to the original phase field simulations; (ii) to propose and compare two sampling strategies, one classical based on Optimal Latin Hypercube Sampling optimized via Particle Swarm Optimization, and one adaptive guided by model uncertainty, to assess whether adaptive sampling reduces the number of required samples while improving performance and overall training efficiency; and (iii) to evaluate two feature extraction approaches, a domain-knowledge-informed transformation coupled with XGBoost and a   convolutional neural network (CNN), to compare generalization capabilities in the absence of domain knowledge against convergence and training cost. Furthermore, all analyses are conducted from a sustainability perspective, quantifying the $CO_2$ emissions associated with each approach.

The remainder of this paper is organized as follows. Section~\ref{sec:methods}, methods section, describes the phase-field modeling framework, data generation process, surrogate modeling approaches, sampling strategies, and experimental setup. Section~\ref{sec:results}, results and discussion section, presents the evaluation of computational cost, $CO_2$ emissions, and predictive performance, together with a discussion of the findings. Finally, Section~\ref{sec:conclusions}, conclusions section, summarizes the main outcomes and implications of the study.

\section{Methods}\label{sec:methods}

\subsection{Phase field model for dendritic solidification in metals}\label{sec:phase-field}

The geometry of the solidification microstructure in metals plays a critical role in manufacturing processes due to its strong influence on the resulting mechanical properties. Among the manufacturing processes in which microstructure is studied are casting, welding, and additive manufacturing. These are typically the initial steps in complex process chains, and the quality of the resulting materials has a direct impact on the final products. In materials science, the complexity of solidification processes is studied across multiple scales, from the atomic to the macroscopic. In general, solidification involves complex interactions between thermal, mechanical, and chemical phenomena.

\textit{Phase field} model is one of the most widely used and effective methodologies for studying, approximating, and simulating materials microstructural evolution. This model is capable of predicting complex, random geometries within continuous field (\cite{tourret_phase-field_2022}). Phase Field methods employ an order parameter, hereafter denoted as $OP$ or $\varphi$, which can take values in the range $\varphi \in [-1, 1]$ or $\varphi \in [0, 1]$. The intermediate values between the interval extremes represent the interface between two distinct physical states of the system: liquid and solid.

\textit{Phase field model} is a numerical method capable of simulating microstructural evolution. As with many other numerical methods, solving the physical differential equations governing solidification processes requires significant computational cost. When large parameter spaces are to be explored, the use of the model can become computationally prohibitive (\cite{alizadeh_managing_2020}).

The model analyzed in this work simulates two-dimensional solidification with dendritic geometries characteristic of metal solidification (\cite{biner_programming_2017}). This simplified model does not account for phenomena resulting from fluid dynamics, phase transition-induced expansion or shrinkage, nor thermal noise. The microstructural evolution is described through the temporal changes of a non-conserved order parameter, as well as spatial variations in temperature across the domain.

Systems evolution is described by the Allen-Cahn equation,

\begin{equation}
\tau \frac{\partial \varphi}{\partial t}=-\frac{\delta F}{\delta \varphi},
\end{equation}

where $\tau$ is the characteristic time scale. The term on the right-hand side of the equation denotes the functional derivative of Ginzburg-Landau free energy, which is defined by, 

\begin{equation}\label{functional}
F(\varphi, m)=\int_V \frac{1}{2} \varepsilon^2|\nabla \varphi|^2+f(\varphi, m) dv,
\end{equation}

where $f(\varphi), m$ is the density of the local free energy. In this work, $f(\varphi, m)$ is defined by a double-well potential, in such a way that it has two stable states or local minima, located at $\varphi=0$ and $\varphi=1$, Thus, $f(\varphi, m)$ is defined in the following way,

\begin{equation} \label{freeEnergyForm}
f(\varphi, m)=\frac{1}{4} \varphi^4-\left(\frac{1}{2}-\frac{1}{3} m\right) \varphi^3+\left(\frac{1}{4}-\frac{1}{2} m\right) \varphi^2.
\end{equation}.

The $\varepsilon$ parameter describes the thickness of the interface. Anisotropy is related to $\varepsilon$ and depends on the the outward-pointing normal vector of the interface. In particular, 

\begin{equation}
\label{epsBarr}
\varepsilon=\bar{\varepsilon} \sigma(\theta),
\end{equation}

where $\bar{\varepsilon}$ defines the mean value of $\varepsilon$ and the anisotropy $\sigma(\theta)$ is expressed as

\begin{equation}
\label{sigma}
\sigma(\theta)=1+\delta \cos \left(j\left(\theta-\theta_o\right)\right).
\end{equation}

Anisotropy is related to the degree of symmetry observed in the resulting dendritic crystals. For example, when $j = 4$, cubic symmetry is obtained, and when $j = 6$, hexagonal symmetry arises. The parameter $\delta$ in Equation~\ref{sigma} defines the strength of the anisotropy. The variable $\theta$ represents the angle relative to a reference direction, as described in Equation~\ref{theta}, while $\theta_0$ denotes the initial orientation angle (\cite{dantzig_solidification_2009}).

\begin{equation}
\label{theta}
\theta=\tan ^{-1}\left(\frac{\partial \varphi / \partial y}{\partial \varphi / \partial x}\right)
\end{equation}

In Equation~\ref{freeEnergyForm}, the parameter $m$ defines the thermodynamic driving force. As the melt becomes undercooled, the well tilts toward the solid phase, causing the order parameter to shift in that direction. The value of $m$ is defined as a function of temperature as follows:

\begin{equation}
\label{mSupercooling}
m(T)=\left(\frac{\alpha}{\pi}\right) \tan ^{-1}\left[\gamma\left(T_{e q}-T\right)\right]
\end{equation}

where $\alpha$ is a positive constant satisfying $\alpha < 1$ (\cite{kobayashi_modeling_1993}). The $T_{eq}$ value of Equation~\ref{mSupercooling} is the equilibrium temperature. Considering the explicit form of the local energy density, it is possible to obtain the evolution of the order parameter in the following way:

\begin{equation}
\label{funcitonalDerivative}
\tau \frac{\partial \varphi}{\partial t}=\frac{\partial}{\partial y}\left(\varepsilon \frac{\partial \varepsilon}{\partial \theta} \frac{\partial \varphi}{\partial x}\right)-\frac{\partial}{\partial x}\left(\varepsilon \frac{\partial \varepsilon}{\partial \theta} \frac{\partial \varphi}{\partial y}\right)+\nabla \cdot\left(\varepsilon^2 \nabla \varphi\right)+\varphi(1-\varphi)\left(\varphi-\frac{1}{2}+m\right).
\end{equation}

Finally, the evolution of the normalized temperature is defined as follows:

\begin{equation}
\label{temperatureEquation}
\frac{\partial T}{\partial t} = \nabla^2 T + \kappa \frac{\partial \varphi}{\partial t}.
\end{equation}

From Equation~\ref{temperatureEquation}, it can be inferred that the equilibrium temperature is 1. The parameter $\kappa$ defines the diffusion constant, which is assumed to have the same value in both the liquid and solid phases.

\subsection{Phase field data}\label{sec:data}

As previously described, the physical model is governed by seven parameters that define the thermophysical conditions of the dendritic solidification. Each specific combination of these parameters leads to a distinct dendritic solidification pattern associated with a given metal under well-defined processing conditions. Table~\ref{tab:physical-params} summarizes the parameters and their corresponding admissible ranges considered in this study.

\begin{table}[htbp]
\centering
\resizebox{\textwidth}{!}{
\begin{tabular}{|c|c|c|c|}
\hline
\textbf{Parameter} & \textbf{Definition} & \textbf{Min value} & \textbf{Max value}\\
\hline
$\tau$     & Characteristic time scale & $1.72\cdotp 10^{-4}$ & $6\cdotp10^{-4}$ \\
$\varepsilon$     & Interface thickness & $0.005$ & $1.95\cdotp10^{-2}$\\
$\kappa$     & Diffusion constant & 1 & 3.6\\
$\delta$ & Related to anisotropy & 0.001 & 0.04\\
j     & Dendritic symmetry & 4 & 6\\
$\alpha$     & Constant related to thermodynamic driving force & 0.4 & 0.9\\
$\theta_0$     & Initial orientation & 0.1 & 0.4\\
\hline
\end{tabular}}
\caption{Minimum and maximum feasible values for each physical model in the phase-field model.}
\label{tab:physical-params}
\end{table}

For each admissible combination of the seven parameters, a full spatio-temporal simulation of the dendritic solidification process is performed. The spatial domain is discretized into a uniform grid of $100 \times 100$ cells, while the temporal evolution is resolved over 4000 discrete time steps. Consequently, at each time step, the phase-field model produces two matrices of size $100 \times 100$: one corresponding to the temperature field and the other to the phase (solidification) field. These outputs can equivalently be interpreted as two-channel images of size $100 \times 100$ pixels.

The objective of the surrogate modeling strategy is to predict the final solidification geometry at $t = 4000$ from partial spatio-temporal information generated by the physical phase-field model. By replacing the full phase-field simulation with a trained artificial intelligence model, a significant reduction in computational cost can be achieved.

However, this task is non-trivial. Let $t_f$ denote the last temporal instance provided as input to the surrogate model. When $t_f$ is closer to $t = 4000$, the prediction accuracy is expected to improve, since the solidification geometry at $t_f$ more closely resembles the final state. Nevertheless, this choice implies that only the computational cost associated with the simulation interval $[t_f, 4000]$ is saved. Conversely, selecting a smaller $t_f$ increases the potential computational savings but makes the prediction task more challenging.

Therefore, the selection of $t_f$ must be carefully balanced. An appropriate choice should ensure sufficient predictive accuracy while maximizing computational savings with respect to the full phase-field simulation. In addition, minimizing the number of training samples required to reach a satisfactory level of accuracy remains a primary objective, as it directly impacts the overall efficiency of the surrogate modeling approach.
\subsection{Surrogate Models}\label{sec:surrogate-models}

As mentioned, when the \textit{Phase field} model is surrogated, the greatest difficulty lies in approximating the spatio-temporal nature of the problem.

Several proposals have been made in the literature to address such problems. For example, the work by (\cite{montes_de_oca_zapiain_accelerating_2021}) extracts the spatial characteristics of the problem using statistical methods and dimensionality reduction techniques. To approximate the temporal relationship between images and predict the final state, it employs an LSTM (Long Short-Term Memory), which is a type of recurrent neural network. The study by (\cite{herman_data-driven_2020}) follows a similar methodology for describing spatial features but then uses polynomial chaos for regression and final state prediction. In contrast, the work by (\cite{peivaste_machine-learning-based_2022}) uses a U-Net neural network to learn spatial features and predict the final state.

The studies by (\cite{li_tutorials_2024}) and (\cite{goswami_transfer_2020}) propose a neural network that incorporates physical information (Physics-Informed Neural Network, or PiNN). Although the former presents the approach for a 1D \textit{Phase field} model, it also explains the necessary modifications to apply it to 2D and 3D problems and how to use spatial data in the neural network. The later, on the other hand, is used to predict fracture geometry and specifically solves a 2D problem.

The work by (\cite{yan_machine_2024}) extracts spatial information using an autoencoder, and then approximates the temporal relationship between images using an LSTM. Similarly, (\cite{vasylenko_element_2021}) also proposes autoencoders to automatically learn the characteristics of the images.


As mentioned in the introduction, other works have proposed neural networks capable of learning both spatial and temporal relationships simultaneously  \cite{ahmad_accelerating_2023}.  Finally, there are also proposals that incorporate information from the differential equations of the \textit{Phase field} model into the neural networks themselves \cite{goswami_transfer_2020}.

One of the main hypotheses of this work is that, although CNN models are capable of learning complex spatial problems without prior information, training such models may require a large amount of data in order to adjust all the model parameters effectively.

However, if prior knowledge about the problem is available, through proper extraction of spatial features, the problem can be approximated using a simpler surrogate model, which requires fewer data to converge.

In a context where data generation is expensive, defining surrogate models that can be efficiently trained is essential. Therefore, the following modeling approaches are proposed:
\begin{enumerate}
    \item \textbf{Spatial Features + XGBoost:}
    The spatial features of the images are extracted using statistical methods. These features are then used to train an XGBoost model, which leverages the spatial characteristics of the input images to predict the final solidification state.
    \item \textbf{Convolutional Neural Network (CNN):}
    The convolutional neural network uses convolutional layers to automatically learn the spatial features of the input images. These learned features are then used to predict the final solidification state.
    \item \textbf{Self-supervised Convolutional Neural Network:}
    In this method, the neural network is pre-trained using a variational autoencoder. The autoencoder learns to reconstruct images based on their features by first distorting the input images used for training. The convolutional layers of the autoencoder are the same as those used in the CNN model. During CNN training, the convolutional layer parameters are initialized with the pre-trained weights from the autoencoder. The goal in this case is to leverage the image representations learned by the autoencoder.
\end{enumerate}

\subsubsection{XGBoost surrogate}\label{sec:XGB}

\begin{figure}
    \centering
    \includegraphics[width=0.85\linewidth]{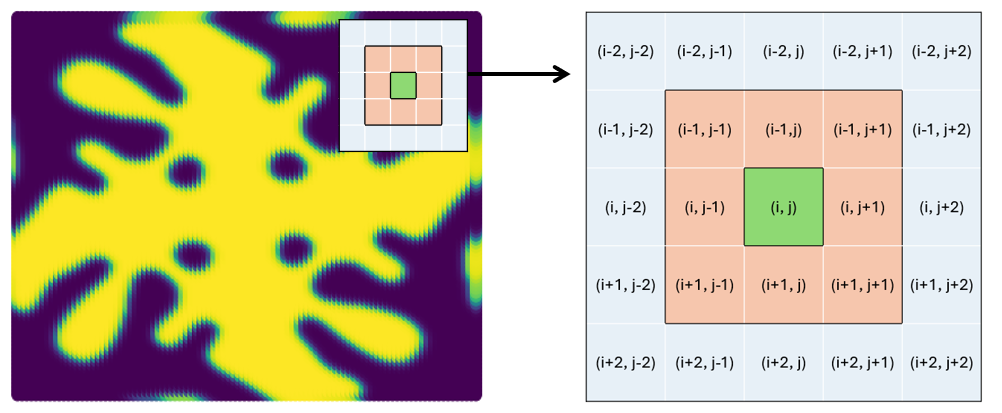}
    \caption{Extracted spatial features for the XGBoost model and for $(i,j)$ pixel.}
    \label{fig:XGB-espazialak}
\end{figure}

\textbf{Spatial features extraction}\\
For each image used to predict the final solidification state, spatial features are extracted. For each cell $(i, j)$ in the image, the spatial features consist of the values of the order parameter and temperature within a radius of $r=2$ around that cell. Figure~\ref{fig:XGB-espazialak} schematically illustrates the extraction of pixel features used for each $(i, j)$ cell.

For each pixel $(i, j)$ and a radius of $r=2$, as shown in Figure~\ref{fig:XGB-espazialak}, the values of temperature and order parameter from the 24 neighboring pixels are used, in addition to the values at $(i, j)$. Therefore, if $n_t$ time steps or images are used to predict the final solidification state, there will be $2 \cdot 25$ spatial features per pixel, resulting in a total of $2 \cdot 25 \cdot n_t$ spatial features for the regression problem.

In addition to the spatial features, the values of the physical simulation parameters from the physical model are also included.\\

\textbf{Symmetry simplification}\\
To reduce the complexity of the regression problem solved using XGBoost, the symmetries of the dendritic geometries are exploited. As explained earlier, this work analyzes dendrites with different anisotropy levels, exhibiting 4-fold and 6-fold symmetry patterns. Following symmetries can be identified:

\begin{itemize}
    \item \textbf{Anisotropy with 4-fold symmetry:}\\
    In this case, the dendritic geometry exhibits symmetry with respect to the axes $x=0$ and $y=0$. Therefore, considering the quadrant defined by the cell range $0 \leq i \leq 50$ and $0 \leq j \leq 50$ as the reference, the full image can be reconstructed by applying rotations of $90^\circ$, $180^\circ$, and $270^\circ$ to this quadrant. Alternatively, if only one axis of symmetry is used, the half-image defined by $0 \leq i \leq 50$ and $0 \leq j \leq 100$ can serve as the reference, and the full image can be recovered by applying a $180^\circ$ rotation.
    \item \textbf{Anisotropy with 6-fold symmetry:}\\
    In this case, the dendritic pattern exhibits rotational symmetry that repeats 6, 3, and 2 times.
    Using the 6-fold rotational symmetry, if the region defined by $0 \leq i \leq 50$ and $0 \leq j \leq \frac{1}{3}i + \frac{100}{3}$ is taken as the reference, the entire image can be reconstructed by applying rotations of $60^\circ$, $120^\circ$, $180^\circ$, $240^\circ$, and $300^\circ$.
    Alternatively, using the 3-fold symmetry and the region $0 \leq i \leq 50$ and $0 \leq j \leq -\frac{1}{3}i + \frac{200}{3}$ as reference, the full image can be reconstructed through $120^\circ$ and $240^\circ$ rotations.
    The 2-fold rotational symmetry case is handled in the same way as the 4-fold anisotropy case.
\end{itemize}

In any case, thanks to the symmetry-based simplification, the model can be trained using only parts of the image. In this way, the final solidification state of a region can be predicted using less information. Then, by applying the corresponding rotations, the full image can be reconstructed.

In the work described in this chapter, the 4-fold rotational symmetry simplification is used for dendrites with anisotropy 4. In contrast, for dendrites with anisotropy 6, the 6-fold rotational symmetry simplification is applied.

\textbf{XGBoost}\\
The XGBoost algorithm combines multiple tree-based estimators using bagging, and refines the model iteratively through boosting by focusing on samples that are more difficult to predict.

In this work, the XGBoost Python library is used. Regarding the algorithm's hyperparameters, default values are used in all cases except for the learning rate, maximum depth, and number of estimators, which are set to $0.01$, $5$, and $200$, respectively. The mean squared error is used as the loss function.

\textbf{Feature Selection}\\
For variable selection, the feature importances provided by the XGBoost model are used. To do this, all variables are ranked from most to least important, the cumulative importance is computed, and the top variables that account for up to 0.98 of the total importance are selected.

\subsubsection{Convolutional Neural Network}

\begin{figure}
    \centering
    \includegraphics[trim=2.3cm 1cm 2.3cm 1.3cm, clip, width=0.99\linewidth]{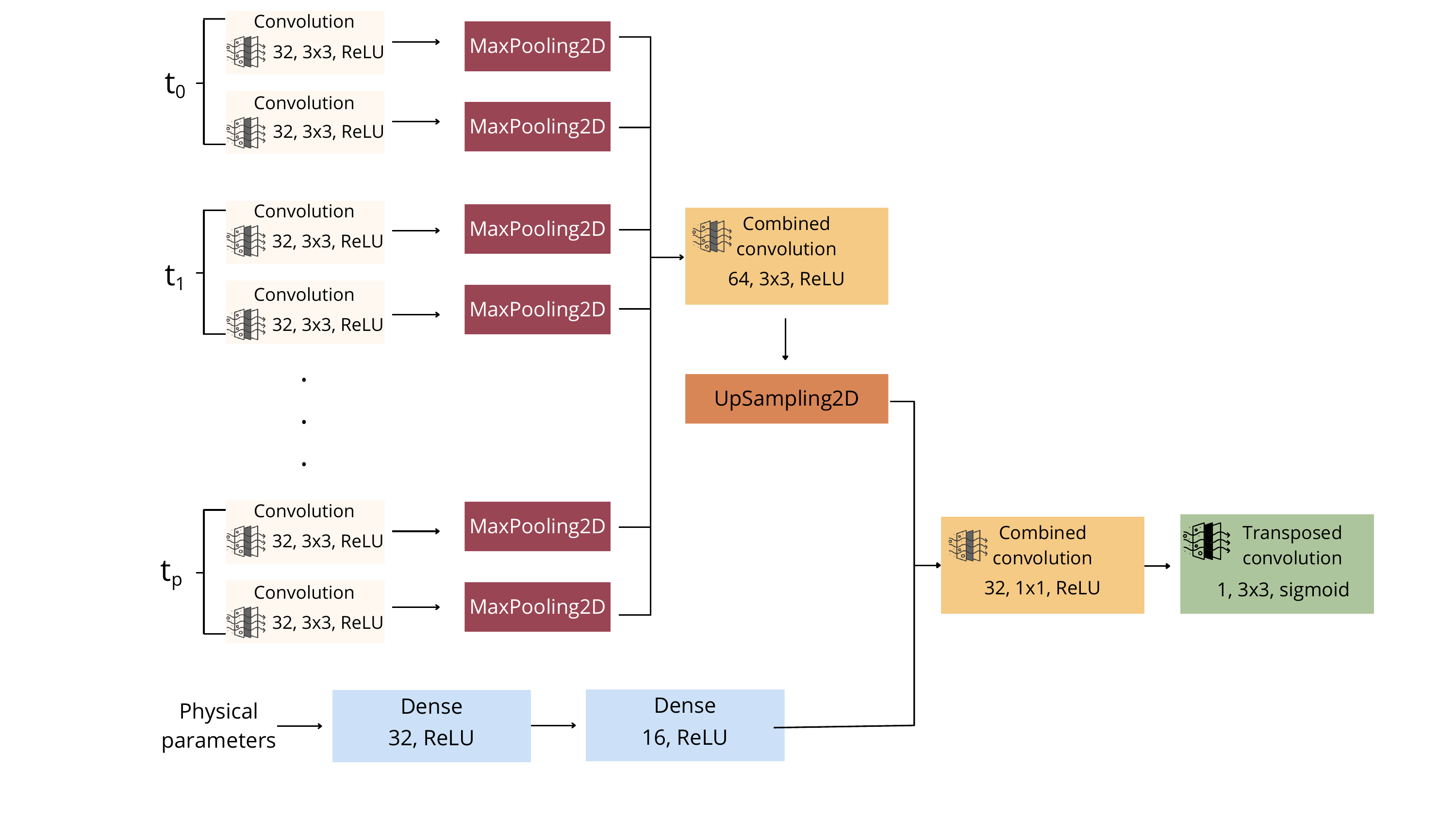}
    \caption{Architecture of the convolutional neuronal network. $t_0$, $t_1$, ..., $t_n$ are the training temporal instances.}
    \label{fig:CNN-egitura}
\end{figure}

The architecture of the convolutional neural network used in this work is summarized in Figure~\ref{fig:CNN-egitura}. The following paragraphs provide a detailed description of each component of the network.

For each training instance and for each sample, one image is generated for temperature values and another for the order parameter. The convolutional neural network applies one convolutional layer to each of these input images. Each of these layers uses 32 filters with kernel sizes of $3 \times 3$, which are applied with a stride of one along both spatial axes. The activation function used is ReLU (Rectified Linear Unit).

Following the convolution, the spatial dimensions of the resulting feature maps are reduced through a downsampling operation that simplifies spatial information. Specifically, max pooling is applied over non-overlapping $2 \times 2$ regions, retaining the maximum value in each case. This reduces the image dimensions by half.

Once all the input images have passed through their respective convolutional layers, their outputs are merged through an additional convolutional layer that combines the extracted features. In this layer, 64 filters are used, with the same kernel size and parameters as in the previous layers. After this operation, the resulting feature map is upsampled to restore the original image dimensions.

The physical parameters used in the simulation, the seven parameters described earlier, are integrated into the neural network. To do this, two fully connected (dense) layers are applied: the first with 32 neurons and the second with 16, both using the ReLU (Rectified Linear Unit) activation function.

The output from the tabular (parameter) layers is then reshaped to match the spatial scale of the image data and combined with the image features through a convolutional layer. This layer uses 32 filters with $1 \times 1$ kernels and again applies the ReLU activation function.

Finally, the prediction of the output image is performed using a transposed convolutional layer followed by a sigmoid activation function. This final layer uses a single filter with a $3 \times 3$ kernel and preserves the spatial dimensions of the input, enabling the network to generate a full-resolution output image.

To train the model, the mean absolute error (MAE) is used as the loss function, and the ADAM optimizer is employed for optimization.

\subsubsection{Self-supervised Convolutional Neural Network}

The self-supervised neural network uses the same convolutional architecture described in the previous section to predict the final solidification state. However, in this case, the convolutional layers of the neural network are pre-trained using a variational autoencoder.

The variational autoencoder adds noise to the input images before processing. Using the same convolutional layers as those in the main network, that is, the layers used initially to process the images, the autoencoder learns to reconstruct the original images from the noisy versions. Through this process, the autoencoder effectively learns to extract meaningful features from the images.

The convolutional layer parameters learned by the variational autoencoder are used to initialize the convolutional layers of the main convolutional neural network. Self-supervised learning is a methodology that has been applied to improve the training of complex neural networks when only few data is available (\cite{mao_self-supervised_2023}). In the case of convolutional neural networks, which typically involve a large number of parameters, the more parameters a model has, the more data it requires to achieve accurate results. This makes properly training convolutional networks particularly challenging (\cite{chen_context_2024}).

When the data originates from a complex physical model, it becomes highly desirable to obtain accurate surrogate models using as little data as possible. In this context, self-supervised learning may provide an effective means to develop more efficient convolutional neural networks, particularly when used as surrogate models.

\begin{figure}
    \centering
    \includegraphics[trim=4.5cm 7cm 4.5cm 7cm, clip, width=0.99\linewidth]{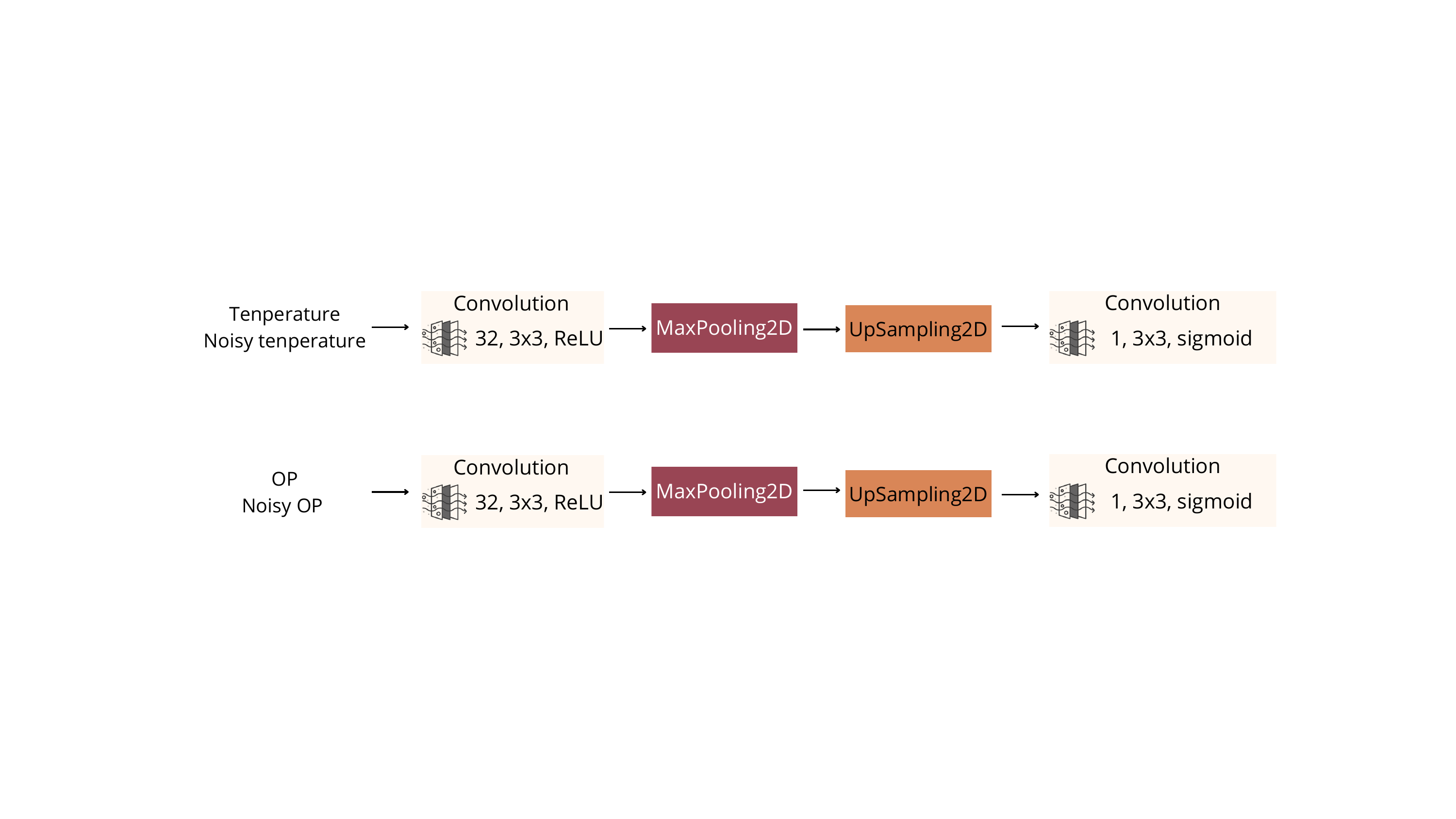}
    \caption{Architecture of the Variational Auto Encoder used in the self-supervised CNN}
    \label{fig:VAE-egitura}
\end{figure}

The architecture of the variational autoencoder used in this work is shown in Figure~\ref{fig:VAE-egitura}. As illustrated, temperature and order parameter images are processed separately through dedicated convolutional layers. These layers are the same ones later used in the main convolutional network, each corresponding to its specific image type. After passing through an encoding–decoding (compression and reconstruction) process, the outputs are further processed by an additional convolutional layer. These final layers use a single filter and apply the sigmoid activation function. The kernel size remains the same as in the initial layers, i.e., $3 \times 3$.

In this way, the autoencoder learns to reconstruct the original images from their noisy versions, effectively capturing the key spatial features of the input images. As mentioned earlier, the parameters obtained after training the first two convolutional layers are then used to initialize the corresponding convolutional layers in the main neural network, as shown in Figure~\ref{fig:CNN-egitura}. Therefore, the convolutional network begins training from a representation that already captures the predicted spatial features.

\subsection{Sampling Methods}\label{sec:sampling}

One of the main objectives of this work is to analyze how the sampling strategy influences the number of samples required to accurately surrogate a computationally expensive problem such as the \textit{Phase field} model, with the ultimate goal of minimizing the number of simulations needed to achieve a good fit.

To this end, two sampling strategies are considered: classical sampling and adaptive sampling. Classical sampling relies on an optimal Latin hypercube design, whereas the adaptive approach incorporates new samples based on model uncertainty. The following sections describe both methodologies in detail.

\subsubsection{Classical sampling: Optimal Latin Hypercube}

Latin Hypercube Sampling (LHS), is one of the most widely used sampling methods in the literature (\cite{luo_review_2023}). The technique divides the space into $m^n$ identical hypercubes to place $m$ points, where $n$ is the dimension of the space. The method consists in distributing the $m$ sample points so that each partition of every coordinate is used exactly once. Specifically, if one of the $m$ points is placed at $(a_1, \dots, a_n)$, and $a_1$ lies in the first cell of the $m$-partition along the first coordinate, then no other point is allowed to share that same cell in the first coordinate. This restriction applies to all remaining coordinates: for each $a_i$, no other point may occupy the same partition cell along that coordinate (\cite{mckay_comparison_1979}).

As is evident, for fixed values of $n$ and $m$, multiple Latin Hypercube Samplings (LHS) can exist. Techniques have been proposed to select the most suitable LHS (\cite{9066677}). While standard LHS ensures a certain level of point dispersion, it does not necessarily guarantee the optimal distribution. For this purpose, various methods have been developed to compute an Optimal Latin Hypercube Sampling (OLHS), which maximizes the distance among all points.

To obtain an OLHS, an optimization problem that maximizes the distance between sample points must be solved. A metric space (most commonly Euclidean) is selected, and an appropriate optimization algorithm is applied. Several methods have been proposed in the literature to generate optimal Latin hypercubes. For instance, (\cite{borisut_adaptive_2023}) reviews different techniques for optimizing LHS, most of which aim to maximize inter-point distances, although some approaches also maximize entropy or incorporate other objective functions.

Classical, metaheuristic, and genetic algorithms have all been employed to solve OLHS problems. For example, in distance-maximization scenarios, \cite{park_efficient_2025} uses an evolutionary algorithm, while \cite{aziz_adaptive_2014} employs a genetic algorithm based on Particle Swarm Optimization (PSO). Other approaches reported in the literature include simulated annealing and various adaptive evolutionary algorithms (\cite{escobar-cuevas_advanced_2024}), as well as propagation–translation LHS methods (\cite{park_efficient_2025}).

In this work, a discrete Particle Swarm Optimization (PSO) approach is proposed to obtain an OLHS. In this optimization problem, each LHS represents a feasible solution. An LHS with $m$ points, in any $n$-dimensional space, can be represented as a permutation of $m$ values along each of the $n$ coordinates. Indeed, in the space $X_1 \times X_2 \times \dots \times X_n$, if a point $(a_1, \dots, a_n)$ is placed, each $a_i$ corresponds to a cell in the LHS along coordinate $i$. Consequently, no other point in the sample may occupy the same LHS cell, i.e., no other point satisfies $X_i = a_i$ for $i = 1, \dots, n$.

Thus, if $m$ uniform cells are defined for each coordinate, each cell contains exactly one point. For instance, consider a two-dimensional case: for the first coordinate, each discretized cell contains a value, which simultaneously corresponds to a coordinate between $1$ and $m$ in the second dimension. The permutation positions indicate the cell in the first coordinate, while the values represent the corresponding positions in the second coordinate. More generally, in an $n$-dimensional sampling, if coordinates are considered in pairs and the sample is represented relative to one another across all coordinates, the LHS can be fully defined by $n$ permutations of $m$ values.

\begin{figure}
    \centering
    \includegraphics[trim=4.5cm 2cm 4.5cm 2cm, clip, width=0.65\linewidth]{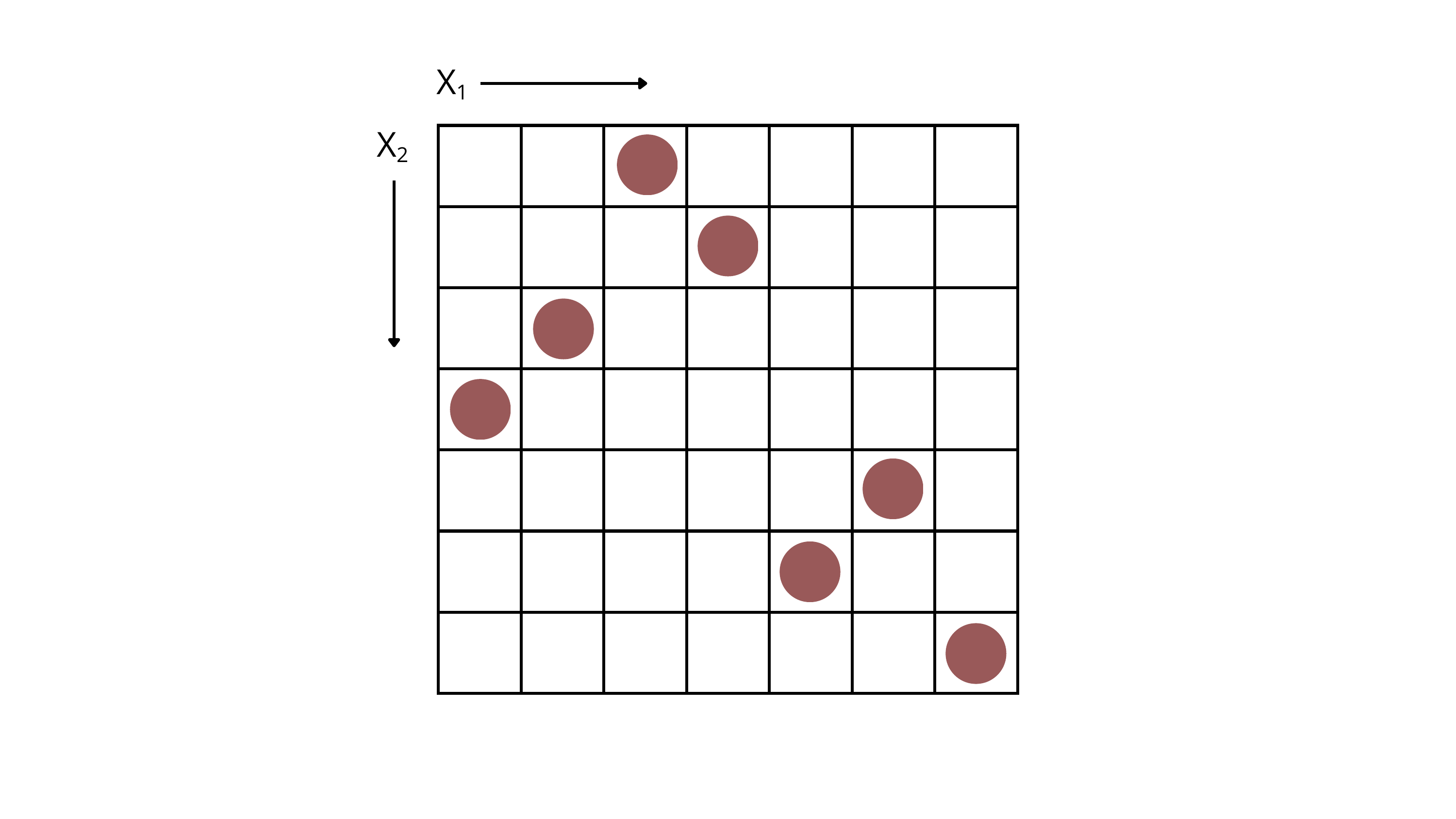}
    \caption{$7$ points LHS sampling for the $[0,1]\times[0,1]$ space. }
    \label{fig:LHS-adibidea}
\end{figure}

As an example, if we want to place 7 points in the $[0,1] \times [0,1]$ space, the LHS illustrated in Figure \ref{fig:LHS-adibidea} can be represented as $\{(4, 3, 1, 2, 6, 5, 7), (3, 4, 2, 1, 6, 5, 7)\}$.

The fitness value of each population is based on the Morris-Mitchell dispersion (\cite{xiao_multi-objective_2022}). A small dispersion value represents a large distance between points. It is calculated as:  

\begin{equation}\label{eq:fitness_LaPSO}
    fitness = \left(\sum_{i = 1}^m \sum_{j = i+1}^m d(p_i, p_j)^{-\delta}\right)^{1/\delta},
\end{equation}

where \(d(p_i, p_j)\) is the distance between the \(i\)-th and \(j\)-th points in the LHS, and \(\delta\) is a pre-defined constant. In this work, the Euclidean distance is used and \(\delta = 50\).

\begin{figure}
    \centering
    \includegraphics[trim=0cm 6cm 0cm 6cm, clip, width=0.98\linewidth]{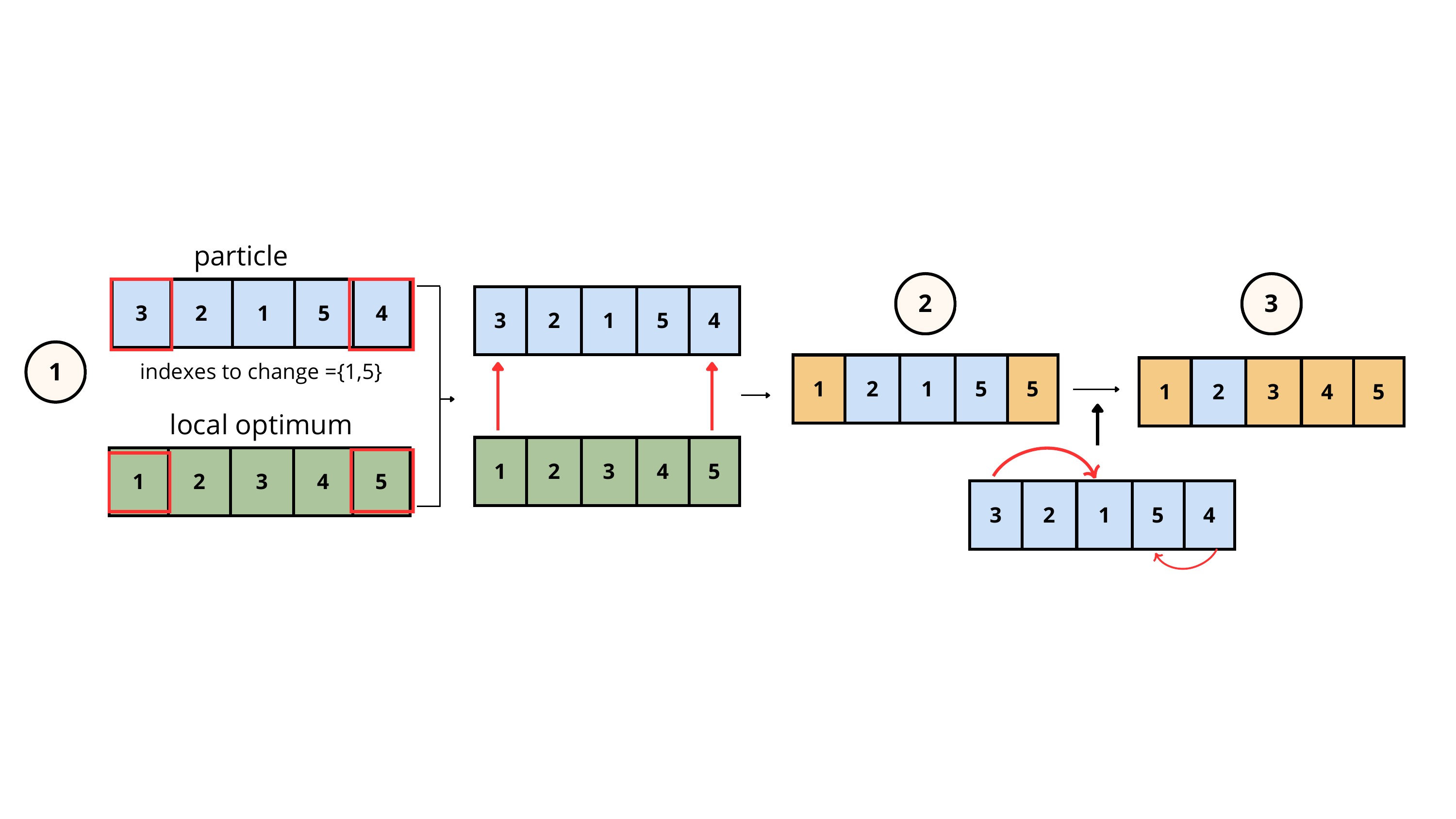}
    \caption{Transformation toward the local and global optimum. The example shows the local optimum case; the global case is symmetric.}
    \label{fig:laPSO-mutazioa1}
\end{figure}

Each population is transformed in every iteration using the following two techniques:

\begin{itemize}
    \item \textbf{Move toward optimum}. The goal is to move the population toward local and global optima. For each parameter \(X_i\), movements are performed toward both the local and global optima. The procedure follows these steps, also illustrated in Figure \ref{fig:laPSO-mutazioa1}:
    \begin{enumerate}
        \item Positions whose values will be randomly modified are selected.
        \item For each value of the first point, the variable is assigned the value of the local or global optimum at that position.
        \item The position where the particle originally held the newly assigned value is identified, and the value at that position is replaced by the value that was just moved.
    \end{enumerate}
    The transformation is first applied toward the local optimum and then toward the global optimum. The number of positions to be changed and the specific positions are randomly selected in each case.
    \item \textbf{Random swap}. Two positions are randomly selected, and their values are exchanged.
\end{itemize}

Movement toward the optima is applied in every iteration, but the movement ``speed,'' i.e., the number of positions changed, varies for each transformation. The random swap is applied probabilistically: for each variable, a value is sampled from a uniform distribution. If this value is smaller than a pre-defined threshold \(R\), a random swap is performed. In this work, \(R = 0.7\) is used.

Finally, the new population is evaluated, and the global and local optima are updated along with the population. The whole algorithm is defined in the pseudocode defined in Algorithm~\ref{alg:LaPSO_LHS}.

\begin{algorithm}
\caption{PSO for Optimal Latin Hypercube Sampling}
\label{alg:LaPSO_LHS}
\begin{algorithmic}[1]
\Require $n\_points, n\_features, n\_pop, max\_iter, tol, distance\_method, p, R$
\Ensure Best LHS design and fitness history

\State Initialize population:
\For{particle $i = 1$ to $n\_pop$}
    \State Generate random LHS of size $(n\_points \times n\_features)$
\EndFor

\State Compute fitness of each particle using Morris-Mitchell dispersion
\State Identify initial global best (gbest) and local best (lbest)
\State Record gbest fitness history

\For{iteration = 1 to max\_iter}
    \If{tol $\ge$ tolerance\_threshold}
        \State \textbf{break}
    \EndIf

    \For{particle $i = 1$ to $n\_pop$}
        \State Extract current LHS$_i$ for particle
        \For{feature $j = 1$ to $n\_features$}
            \State Randomly select number of positions to move toward lbest and gbest
            \State Select indices of positions: same\_num\_p, same\_num\_g
            \State Change particle
            \State new\_LHS$_i\gets$ move\_design(feature$_j$, LHS$_i$, lbest, same\_num\_p)
            \State new\_LHS$_i\gets$ move\_design(feature$_j$, LHS$_i$, gbest, same\_num\_g)
            \If{rand() $<$ R}
                \State new\_LHS$_i$ $\gets$ rand\_swap(LHS$_i$, feature$_j$)
            \EndIf
        \EndFor
    \EndFor

    \State new\_fitness $\gets$ compute\_fitness(new\_population, distance\_method, p, R)

    \For{particle $i = 1$ to $n\_pop$}
        \If{new\_fitness $>$ previous\_fitness}
            \State LHS$_i\gets$ new\_LHS$_i$
            \State fitness$_i\gets$ new\_fitness$_i$
            \If{new\_fitness $>$ gbest fitness}
                \State Update gbest
                \State Record gbest fitness
                \State Update tolerance
            \EndIf
        \EndIf
    \EndFor

    \State Update lbest to particle with maximum fitness in current population
\EndFor

\State \Return gbest LHS and fitness history
\end{algorithmic}
\end{algorithm}

\subsubsection{Uncertainty based adaptive sampling}\label{sec:adaptive-sampling}

Adaptive sampling aims to iteratively improve surrogate models or optimization processes by selecting new samples based on model performance and uncertainty. Classical sampling strategies typically emphasize exploration only, which can be inefficient when the relationship between input variables and the output is highly nonlinear and unknown. Balancing exploration and exploitation is therefore essential for constructing accurate and efficient adaptive sampling (\cite{viana_surrogate_2021}). In surrogate-based modeling, adaptive sampling allows extracting information effectively from computationally expensive simulations as the studied in this work (\cite{viana_surrogate_2021}). As the relationship between the input space and the response is initially unknown, and because the number of samples strongly influences performance, strategies that adaptively refine the design are crucial (\cite{lookman_active_2019}).

Adaptive sampling methods address these challenges by iteratively enriching the training dataset with points that improve the model, striking a balance between exploration and exploitation (\cite{lookman_active_2019, fuhg_state---art_2021}). Among the most effective strategies are those based on epistemic uncertainty estimation. These methods quantify the predictive uncertainty of the surrogate model over the entire design space and add samples in regions where uncertainty is highest. High-uncertainty regions can arise either from sparse sampling (lack of exploration) or from areas with steep response gradients (need for exploitation), even if nearby points exist (\cite{hullermeier_aleatoric_2021}).

Numerous approaches for uncertainty estimation have been proposed in the literature (\cite{heid_characterizing_2023, fakour_structured_2024}), and their effectiveness is closely tied to the surrogate model employed. Probabilistic models offer a more direct way to quantify uncertainty compared to deterministic ones. Nevertheless, reliable uncertainty measures for deterministic surrogates also exist. For neural networks, for example, Monte Carlo dropout provides a practical Bayesian approximation by randomly deactivating neurons during inference (\cite{gal_dropout_2015}). Other model-agnostic approaches include quantile-based loss functions, numerical methods to estimate joint conditional distributions (\cite{tyralis_review_2024}), or model ensemble methods (\cite{NEURIPS2021_a70dc404, hoffmann_uncertainty_2021}). 

Building on these ideas, the adaptive method proposed in this work combines exploration and exploitation through uncertainty-driven sampling. By iteratively measuring and leveraging the model's uncertainty, new points are added in the most informative regions of the design space, enhancing both accuracy and efficiency.

\subsubsubsection{Monte Carlo Dropout for uncertainty estimation}

In the case of neural networks, the \textit{Monte Carlo dropout} method was employed to quantify uncertainty. This approach has been widely adopted to estimate epistemic uncertainty in neural networks (\cite{padarian_assessing_2022, liu_quantitative_2025, zeevi_monte-carlo_2024}). Moreover, as explained in (\cite{seoh_qualitative_2020}), neural networks trained and evaluated with \textit{Monte Carlo dropout} are analogous to Bayesian neural networks and provide reliable uncertainty estimates.

\textit{Dropout} is a regularization technique originally designed to prevent overfitting in neural networks. The method randomly deactivates a proportion of neurons and can be applied during both training and inference. In this work, dropout is applied at two stages during training: first, after each layer that processes the image data (following the structure shown on the left side of Figure~\ref{fig:CNN-egitura}); and second, in the convolutional layer where physical parameter features and image features are combined (the second layer from the right in Figure~\ref{fig:CNN-egitura}).

After training, dropout is also applied during inference, where random network connections are deactivated and ignored for prediction. For each input image, $100$ stochastic forward passes are performed using dropout, resulting in $100$ predictions. The mean of these predictions is taken as the final network output, while the standard deviation across the $100$ predictions serves as the measure of predictive uncertainty.

To obtain a spatial distribution of the model uncertainty and an overall uncertainty estimate for each model, cross-validation is employed. All available samples are divided into $10$ groups of equal size, and during $10$ iterations, one group is used for validation while the remaining $9$ groups are used for training. In each case, uncertainty is computed on the validation samples.

\subsubsubsection{Bagging XGBoost Trees for uncertainty estimation}

For XGBoost, the methodology used combines bootstrap and bagging techniques to estimate the model’s epistemic uncertainty. Similar approaches have been described in the literature for quantifying uncertainty in tree-based machine learning models (\cite{brophy_instance-based_2022, malinin2021uncertainty}). Although the variance induced by bootstrap resampling is not a purely epistemic measure—because subsampling of instances and features introduces additional sampling noise—it has been widely adopted as a practical approximation to epistemic uncertainty in complex regressors, as demonstrated for example in \cite{lookman_active_2019}, \cite{fakour_structured_2024}, and \cite{heid_characterizing_2023}. Moreover, \cite{palmer_calibration_2022} shows that while bootstrap-based uncertainty is not an exact estimator of epistemic error, it remains strongly and monotonically correlated with the true prediction error, supporting its use as a reliable and informative approximation.

As mentioned earlier, the XGBoost implementation employed in this work uses $200$ estimators. Each estimator is trained on a random subset of both features and samples. During inference, a random $\%75$ of the trained estimators is selected for each prediction, and the prediction process is repeated $100$ times per input sample. The mean value of these repetitions is taken as the final prediction, while the standard deviation across the $100$ repetitions serves as the uncertainty measure.

Similar to the methodology adopted for neural networks, cross-validation is also employed here to approximate the overall uncertainty across the entire input space. In this case, $\%20$ of the samples are randomly selected for validation, and $10$ repetitions are performed. In each repetition, uncertainty is computed on the validation samples.

\subsubsubsection{Definition of new samples}

Using the uncertainty estimates obtained for each model, regions of high uncertainty are identified in order to introduce new samples within those regions.

As previously mentioned, uncertainty is computed on the validation samples using the two methodologies described above. In any case, this approach only provides uncertainty values for samples that are already part of the design set. However, the use of cross-validation allows representing uncertainty across the entire input space, particularly considering that the initial sampling is performed via OLHS.

In this work, if the number of samples at iteration $k$ is denoted as $n_k$, then for iteration $k+1$, an additional $\lfloor 0.15 n_k \rfloor$ samples are added to the design set, resulting in $n_{k+1} = \lfloor 1.15 n_k \rfloor$ total samples. At each iteration, these $\lfloor 0.15 n_k \rfloor$ new samples are selected based on the model uncertainty as follows:

\begin{enumerate}
    \item At iteration $k+1$, the $\lfloor 0.15 n_k \rfloor$ candidate points with the highest uncertainty are identified.
    \item For each of these points, the distance to all other points in the design set is computed, and the nearest neighbor is identified.
    \item Hyperspheres are defined with centers at the high-uncertainty points and radius equal to half the distance computed in step 2.
    \item For each hypersphere, one new sample is randomly generated within it. This set of points constitutes the $\lfloor 0.15 n_k \rfloor$ new samples added to the design set.
\end{enumerate}

\begin{figure}[h!]
    \centering
    \includegraphics[trim=2cm 7cm 2cm 7cm, clip, width=0.98\linewidth]{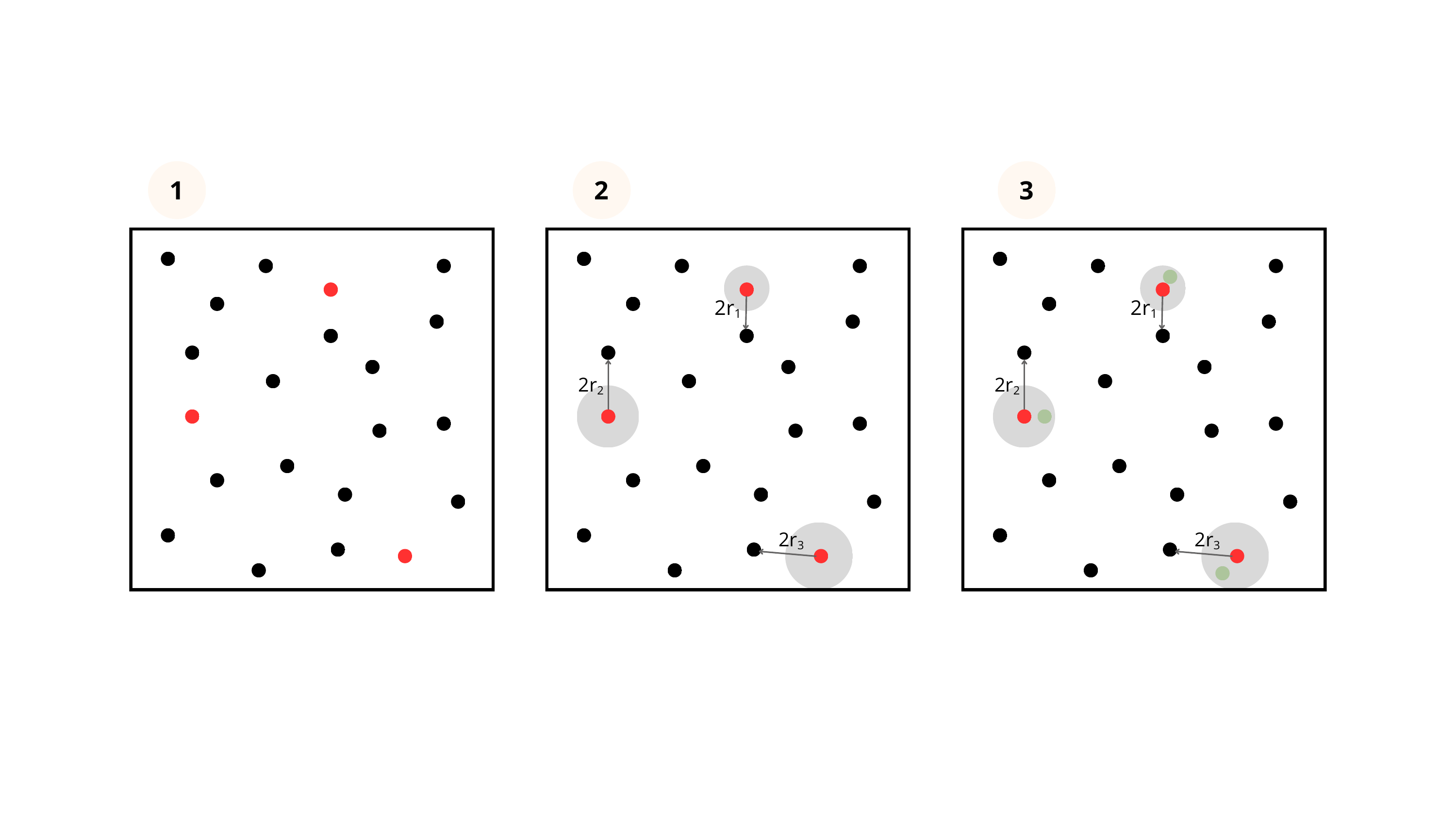}
    \caption{Definition of new samples based on uncertainty estimation. Red points denote the samples with the highest uncertainty, while green points represent those added for iteration $k+1$.}
    \label{fig:adaptive-eskema}
\end{figure}

As an illustrative example, Figure~\ref{fig:adaptive-eskema} depicts the adaptive sampling procedure for two physical parameters at iteration $k$.

\subsubsubsection{Adaptive Sampling Based on Mixed Uncertainty}

The predictive uncertainty of each sample is computed from the pixel-wise uncertainty of its predicted solidification image, using the Monte Carlo Dropout and bagging strategies explained above.

Each sample or image is partitioned into three zones according to the solidification phase predicted for each pixel: pixels with values below $\delta$ are assigned to the \emph{liquid} zone, those above $1-\delta$ to the \emph{solid} zone, and the remaining pixels to the \emph{interface} zone (the transition between liquid and solid), where $\delta$ is a small positive value close to $0$.

For each sample $i$, the mean uncertainty is calculated separately over each zone, denoted by $\bar{U}^{(i)}_{\mathrm{liq}}$, $\bar{U}^{(i)}_{\mathrm{int}}$, and $\bar{U}^{(i)}_{\mathrm{sol}}$. 

The interface region, being the most difficult to predict and determining the final geometry of the solidified metal, and consequently its mechanical properties, is assigned the largest weight when aggregating the uncertainty. 
The mixed uncertainty is therefore defined as:
\[
U^{(i)}_{\text{mixed}} = w_{\mathrm{liq}}\,\bar{U}^{(i)}_{\mathrm{liq}} + w_{\mathrm{int}}\,\bar{U}^{(i)}_{\mathrm{int}} + w_{\mathrm{sol}}\,\bar{U}^{(i)}_{\mathrm{sol}},
\]
with $w_{\mathrm{liq}}=0.1$, $w_{\mathrm{int}}=0.6$, and $w_{\mathrm{sol}}=0.3$. 

These mixed uncertainty values are used within the adaptive sampling algorithm (Algorithm~\ref{alg:adaptive-sampling-uncertainty}) to rank the samples, select those with the highest uncertainty, and generate new candidate points to enrich the design space.

\begin{algorithm}[H]
\caption{Adaptive Sampling Based on Mixed Uncertainty}
\label{alg:adaptive-sampling-uncertainty}
\begin{algorithmic}[1]
\Require $p, uncertainty\_df, doe, w_{\mathrm{liq}}, w_{\mathrm{int}}, w_{\mathrm{sol}}, \delta$
\Ensure New sample points and updated DOE

\State Compute mixed uncertainty for all samples:
\For{each sample $i$ in uncertainty\_df}
    \State Partition predicted solidification image into zones:
        \Statex \quad liquid: pixels $\le \delta$, 
        \Statex \quad solid: pixels $\ge 1-\delta$, 
        \Statex \quad interface: remaining pixels
    \State Compute mean uncertainty per zone: $\bar{U}^{(i)}_{\mathrm{liq}}$, $\bar{U}^{(i)}_{\mathrm{int}}$, $\bar{U}^{(i)}_{\mathrm{sol}}$
    \State Compute mixed uncertainty: 
    \[
        U^{(i)}_{\text{mixed}} = w_{\mathrm{liq}}\,\bar{U}^{(i)}_{\mathrm{liq}} + w_{\mathrm{int}}\,\bar{U}^{(i)}_{\mathrm{int}} + w_{\mathrm{sol}}\,\bar{U}^{(i)}_{\mathrm{sol}}
    \]
\EndFor

\State Sort samples by $U^{(i)}_{\text{mixed}}$ in descending order
\State Select top $p$ most uncertain samples: $p\_most\_uncertain$

\For{each sample in $p\_most\_uncertain$}
    \State Compute scaled coordinates within [0,1] range for all features
    \State Find nearest existing sample and compute half-distance
    \State Define hypersphere centered at sample with radius = half-distance
    \State Generate new candidate point randomly inside hypersphere
    \State Rescale new point back to original feature space
\EndFor

\State Assign new experimental IDs to generated points
\State Concatenate new points with existing DOE
\State \Return new sample points, updated DOE
\end{algorithmic}
\end{algorithm}

\subsection{Experimental setup}\label{sec:experimental-setup}

This section describes the experimental framework developed to evaluate the hypotheses presented in this work.

\subsubsection{Training Temporal Instances}

For spatio-temporal problems such as phase field simulations, it is essential to capture both spatial and temporal features of the system. Spatial features are handled as described in Section~\ref{sec:XGB}. Temporal features require careful selection of discrete time instances used to train the surrogate model. The final state prediction is strongly influenced by the temporal instances closest to the target time.  

The computational cost of surrogate model predictions depends on the number and choice of temporal instances simulated using the original high-fidelity model. Let $t_f$ denote the last time instance used for training; the potential computational savings are inversely proportional to the cost of simulating the instances between the initial time and $t_f$. Based on this principle, three temporal instance sets are defined for evaluation:

\begin{itemize}
    \item $ID_1 = \{1000, 1500, 2000, 2500\}$
    \item $ID_2 = \{1000, 1500, 2000, 2500, 3000\}$
    \item $ID_3 = \{1000, 1500, 2000, 2500, 3000, 3500\}$
\end{itemize}

These configurations correspond to executing approximately half, three-quarters, and seven-eighths of the full simulations required to predict $t = 4000$ using the surrogate.

\subsubsection{Sampling Size}

For adaptive sampling, the initial design follows the recommendations of \cite{tenne_initial_2015, fuhg_adaptive_2019}, and \cite{ afzal_effects_2017}, setting the initial sample size to $10n$, where $n=7$ is the number of physical parameters, yielding $70$ initial points. At each iteration $k+1$, additional samples are added according to Section~\ref{sec:adaptive-sampling} as $\lfloor 0.15 n_k \rfloor$, resulting in a progressive growth of the sample set: $\{70, 80, 92, 105, 120, 138,\dots\}$.

Classical fixed designs do not allow incremental adjustment based on model uncertainty. For comparison, fixed designs of $70, 150, 300, 500,$ and $700$ points are evaluated, with $700$ representing the maximum sample size for both adaptive and classical strategies.

\subsubsection{Incremental Neural Network Training}

Adaptive sampling is coupled with incremental surrogate training. Starting from the initial design, the model is retrained as new samples are added, reducing overall model uncertainty progressively. Incremental training is implemented using the \textit{Keras} library (\cite{chollet2015keras}), which allows updating model weights with new samples while preserving previously learned information.  

Cross-validation is applied to estimate uncertainty across the entire sample space. To avoid biasing uncertainty estimation toward newly added points, incremental retraining is reserved for the final model. This approach maintains both computational efficiency and accurate representation of global uncertainty.

Overall, this experimental design systematically investigates the influence of temporal instance selection, sample size, and adaptive incremental training on surrogate model performance, directly addressing this work's hypotheses.

\subsubsection{Carbon Footprint Analysis}

The accelerating impacts of climate change and global environmental degradation have led to international commitments such as the Paris Climate Change Agreement (PCCA), through which almost all countries have agreed to reduce their carbon emissions (\cite{wang2024ecological}). In this broader context, the rapid expansion of AI has garnered attention due to its substantial energy demands. Despite the potential of AI technologies to optimise energy use and promote environmental sustainability, their development and deployment are themselves energy-intensive processes that generate substantial greenhouse gas emissions.

The amount of computational power used for AI increased by 300,000 times between 2012 and 2018 (\cite{freitag2021real}). Consequently, researchers have emphasised the significance of adopting a "Green AI" approach, which prioritises computational efficiency and sustainability, as a contrasting approach to the current prevailing "Red AI" approach, which prioritises model accuracy without considering environmental cost (\cite{freitag2021real}). This has led to the integration of monitoring tools to quantify and mitigate the energy consumption and carbon footprint of AI workloads.

This study employs CodeCarbon (\cite{benoit_courty_2024_11171501}), an open-source software, to estimate $CO_2$ emissions and energy consumption generated by training tasks. CodeCarbon records energy consumption from CPUs, GPUs, and memory, and converts these data into carbon-equivalent values based on the carbon intensity of the geographical region. This tool serves to complement traditional performance indicators, such as the $R^2$, with quantitative metrics that reflect environmental costs.

\section{Results and discussion}\label{sec:results}
	
To evaluate and validate the hypotheses and proposed methodology, the results are analyzed along three main dimensions: (1) the environmental impact of each strategy, quantified through its carbon footprint at the same steps; (2) computational costs associated with simulation, sampling, and training; and (3) model performance metrics achieved with each approach. Finally, these dimensions are jointly assessed to highlight the overall efficiency and sustainability attainable under each scenario.

In the following sections, the results are presented using the following notation for each model: XGB refers to the XGBoost model trained with classical sampling, and XGBoost\_stochastic denotes the same model trained with adaptive sampling; CNN refers to the convolutional neural network trained with classical sampling, and CNN\_MC\_dropout indicates its adaptive sampling counterpart; finally, CNN\_ss denotes the self-supervised convolutional neural network trained with classical sampling, and CNN\_MC\_dropout\_ss refers to its adaptive sampling version.

\subsection{$CO_2$ emissions}\label{sec:CO2-emissions}

The metrics recorded with CodeCarbon during the training sessions provided a detailed account of both $CO_2$ emissions and energy consumption (kWh). The server was deployed in Sweden, and CodeCarbon automatically applied the Swedish energy data to calculate emissions and energy usage. Sweden was selected because it has one of the lowest emission intensities for electricity generation in Europe, as published in the report presented by the European Environment Agency (EEA) \cite{noauthor_greenhouse_2025}. 

\begin{figure}[htbp]
    \centering
    \begin{subfigure}[b]{0.49\textwidth} 
        \centering
        \includegraphics[width=\textwidth]{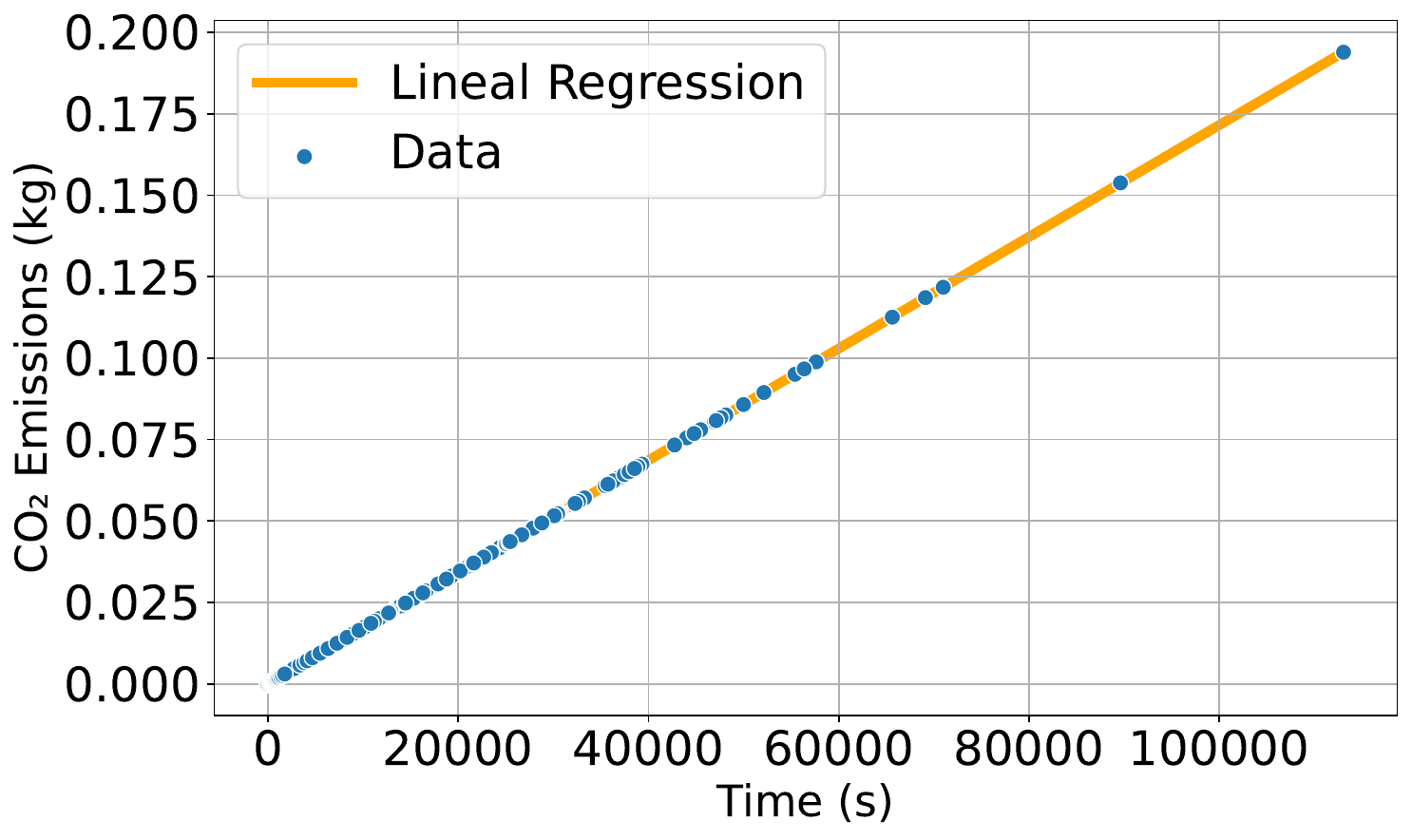}
        \caption{$CO_2$ emissions (kg) over time (s)\\$\hat{y} = 0.00000172x + -0.00000013$}
        \label{emissions_a}
    \end{subfigure}
    \hfill
    \begin{subfigure}[b]{0.49\textwidth}
        \centering
        \includegraphics[width=\textwidth]{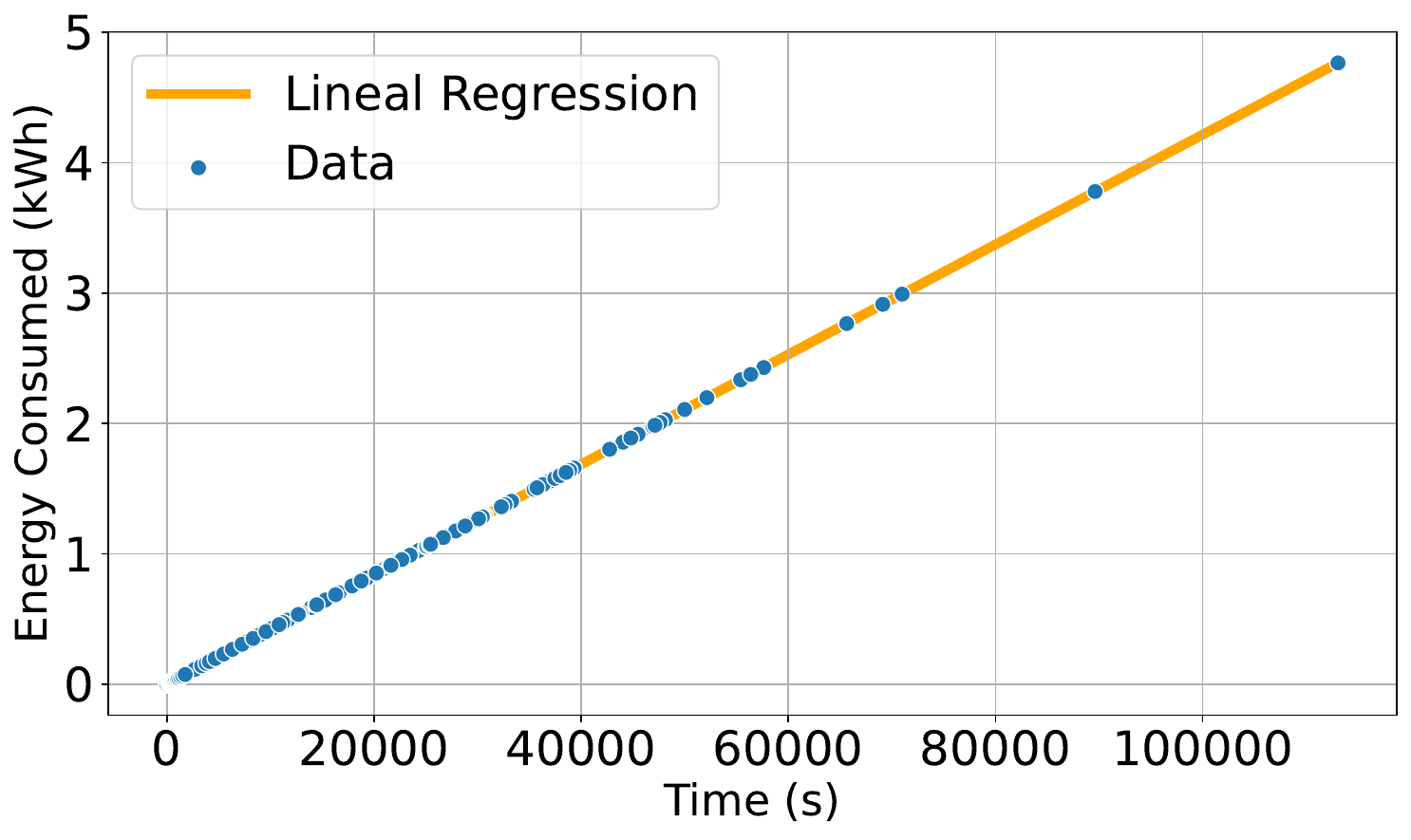}
        \caption{Energy consumed (kWh) over time (s)\\$\hat{y} = 0.00004215x + -0.00000309$}
        \label{emissions_b}
    \end{subfigure}

    \caption{CodeCarbon metrics with respect to duration}
    \label{fig:codecarbon_results}
\end{figure}

Considering all the experiments carried out during this work, Figure \ref{fig:codecarbon_results} shows the relations between the $CO_2$ emissions and energy consumption vs experiments computational temporal costs, subfigure \ref{emissions_a} and \ref{emissions_b} respectively.

The visible linear correlation is supported by a linear regression model, where the coefficient of determination ($R^2$) approaches 1 (0.99999997 for both $CO_2$ emissions and energy consumption). This indicates that time, according to CodeCarbon, is the primary factor influencing both emissions and energy use during the training processes. Consequently, the following sections will directly evaluate the environmental impact of the experiments in respect of time, as this is a reliable indicator of the overall impact.

\subsection{Computational cost}\label{sec:comp-costs}

When assessing the computational cost required to fit a surrogate model, it is necessary to measure the execution time of each step in the process. Specifically, the fitting of a surrogate model involves three main stages: (1) data generation through simulations, (2) sample selection using a sampling method, and (3) model training.

The methodology proposed in this work focuses on the sampling strategies and the models themselves. Therefore, in order to enable a fair comparison across all experiments, the computational costs analyzed correspond to stages (1), (2) and (3).

Regarding data generation, in this case all simulations are performed with the same physical model. Consequently, the computational cost of generating simulated samples is constant, disregarding minor fluctuations due to background processes of the computing system.

However, the scope of the surrogate model strongly depends on both the quantity and the quality of the available data. For this reason, it is still essential to consider the number of samples, although this aspect will be analyzed in a later section, as the objective is to maximize model performance with the minimum amount of data. Importantly, in the context of more complex physical models, such as full \textit{Phase field} simulations, the computational cost of data generation becomes a critical factor, where both the number of samples and the total simulation time play a decisive role.

The following subsections analyze the computational costs associated with each of the steps outlined above.

\subsubsection{Simulations temporal cost}

\begin{figure}
    \centering
    \includegraphics[width=0.75\linewidth]{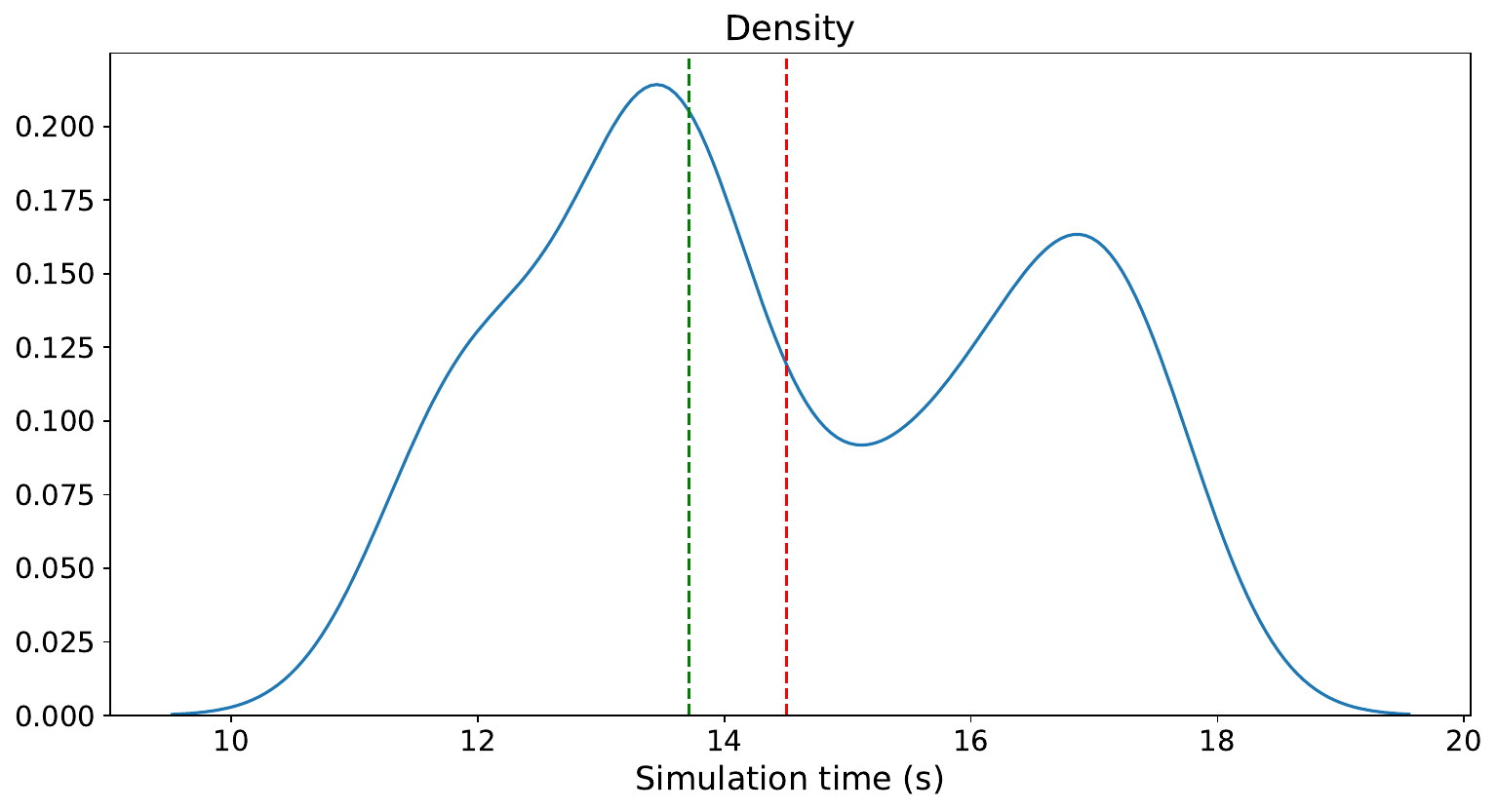}
    \caption{Density distribution of the time required to simulate a single sample by the phase field model.}
    \label{fig:sim-histo}
\end{figure}

To approximate the simulation time of the original physical model, the computational cost of all simulations was recorded. Figure \ref{fig:sim-histo} shows the density distribution of the simulation times, where the vertical lines indicate the mean (red) and the median (green), which are $14.5s$ and $13.7s$, respectively.

\subsubsection{Sampling temporal cost}

As mentioned above, two sampling methods were considered: classical and adaptive. For the classical method, the computational cost corresponds to the time required to generate the samples for a given sample size, namely $\{70, 150, 300, 500, 700\}$.

In the case of adaptive sampling, the computational cost for a sample size $n_k$ must also include the accumulated cost of all previous iterations, i.e., $n_0,...,n_{k-1}$. This is because reaching $n_k$ samples requires $k$ iterations of the method. Thus, in addition to the initial classical sampling, the adaptive method is executed $k$ times.

\begin{figure}[htbp]
    \centering

    \begin{subfigure}[b]{\textwidth}
        \centering
        \includegraphics[width=0.8\textwidth]{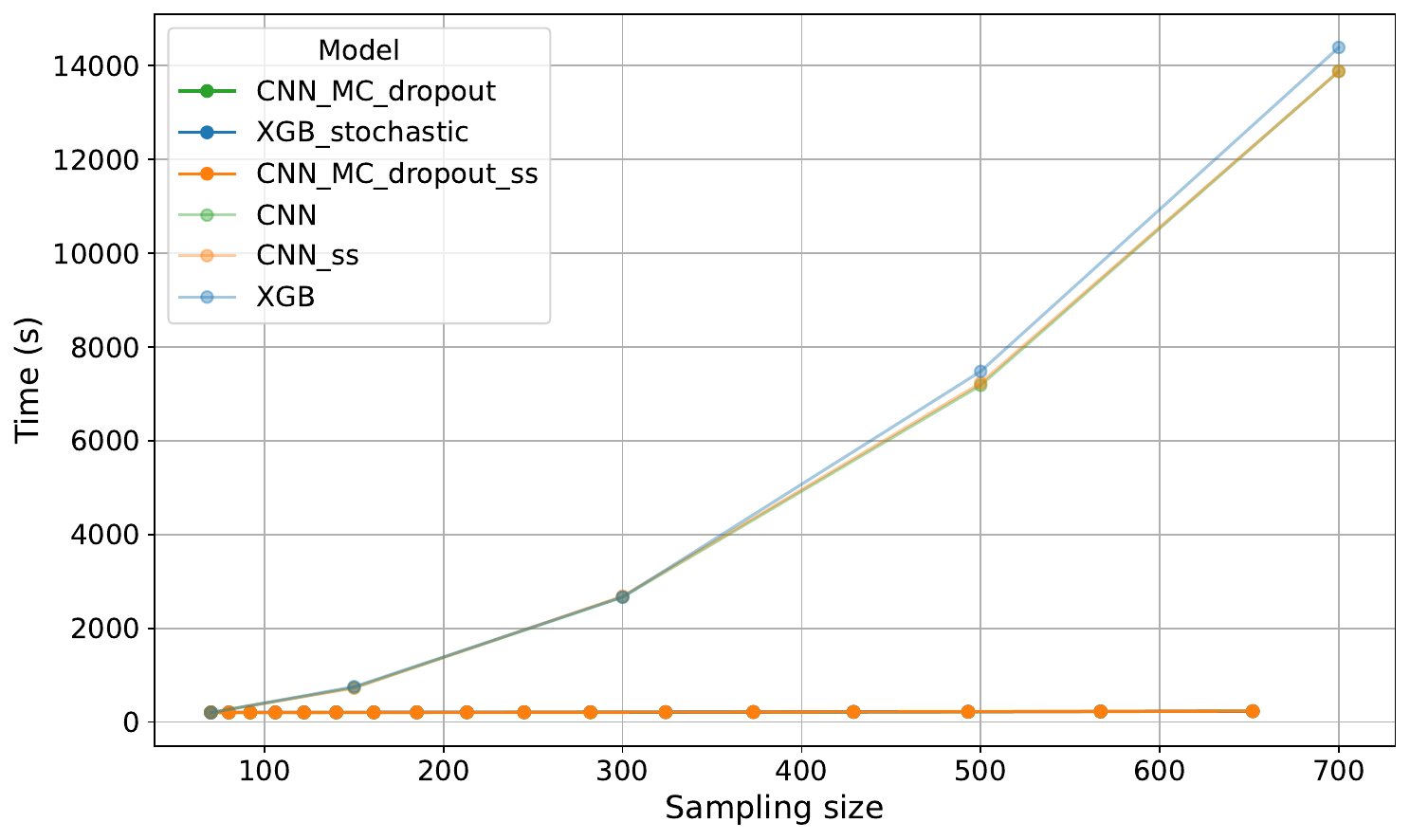}
        \caption{Sampling cost for training instance ID1.}
        \label{fig:sub1}
    \end{subfigure}
    \hfill
    \begin{subfigure}[b]{\textwidth}
        \centering       
        \includegraphics[width=0.8\textwidth]{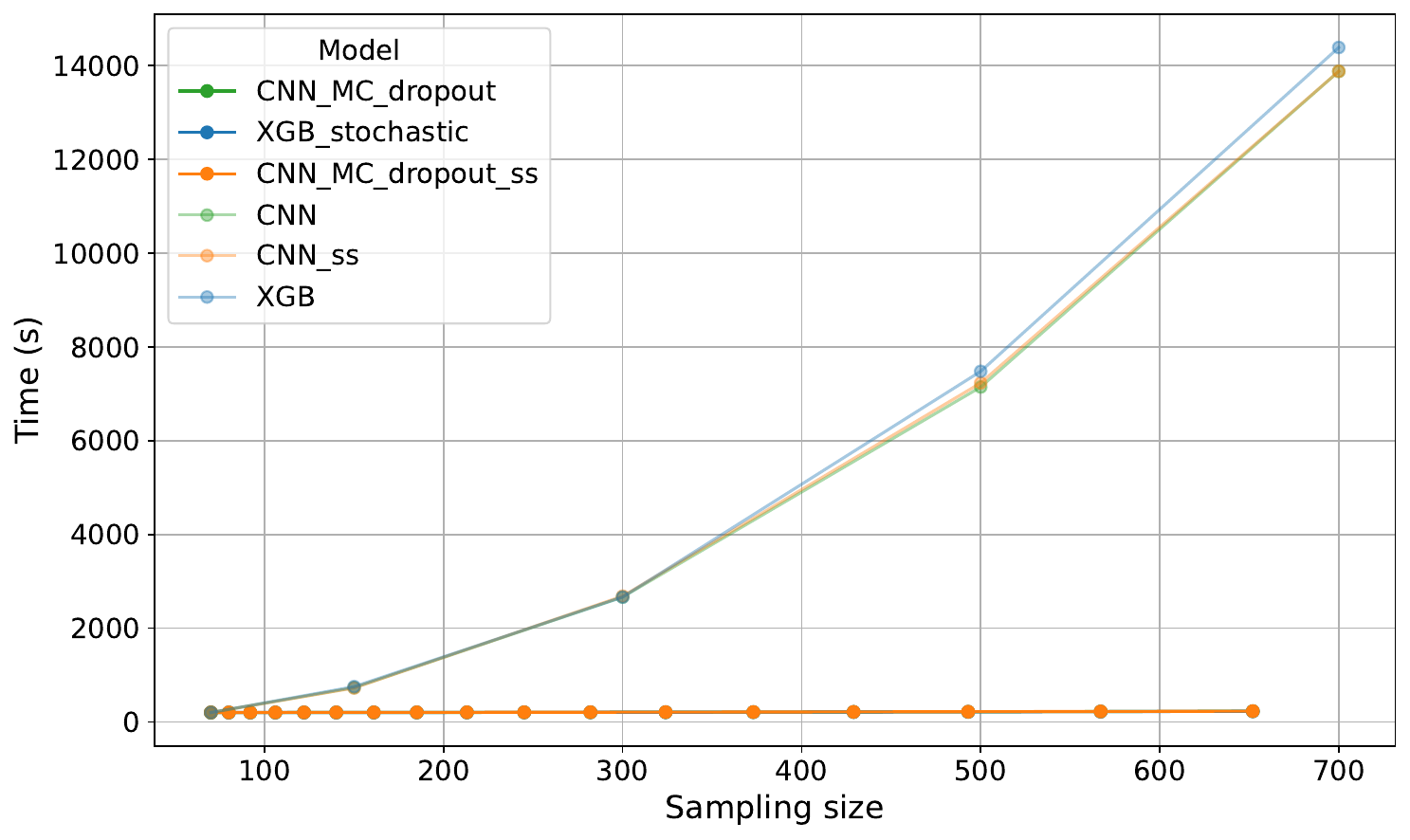}
        \caption{Sampling cost for training instance ID2.}
        \label{fig:sampli}
    \end{subfigure}
    \hfill
    \begin{subfigure}[b]{\textwidth}
        \centering
        \includegraphics[width=0.8\textwidth]{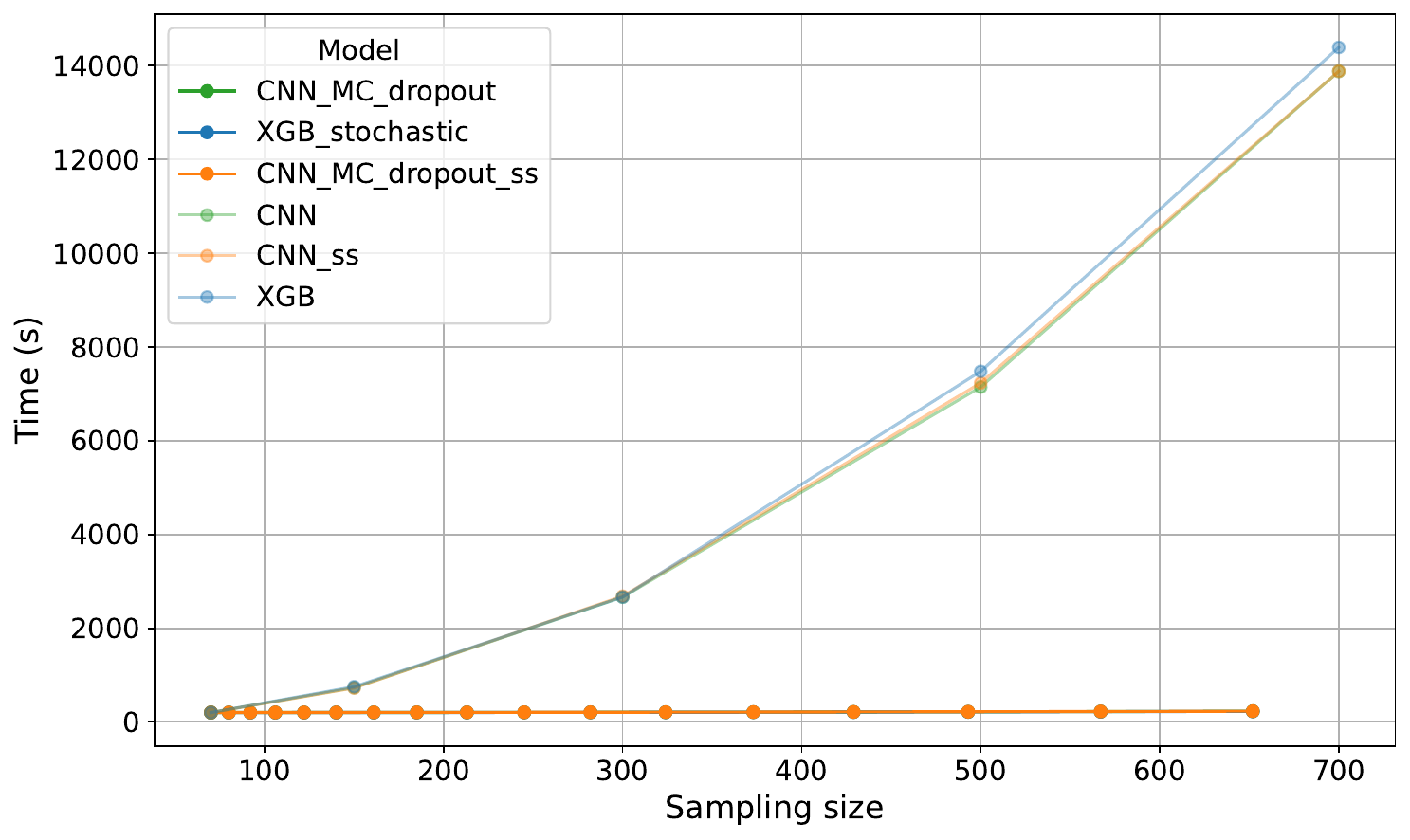}
        \caption{Sampling cost for training instance ID3.}
        \label{fig:sub3}
    \end{subfigure}

    \caption{Computational cost of sampling for each group of training instances, across all models and both sampling methods.}
    \label{fig:laginketa-kostea}
\end{figure}

Figure \ref{fig:laginketa-kostea} presents the results for the three experimental instances. Each plot shows the computational cost (time) for each model and sampling size. The same colors are used for each type of model, while lighter shades indicate the cost of the classical sampling methodology, in contrast with the adaptive method.

The number of training instances and the images used do not influence the sampling method, since the latter only selects which parameters of the physical model should be used to generate new samples in order to improve the surrogate model uncertainty. As shown in Figure \ref{fig:laginketa-kostea}, the cost of each method depends solely on the sample size, with no impact from the model itself. Moreover, the results clearly indicate that adaptive sampling entails significantly lower computational costs compared to classical sampling.

The difference between the two strategies is noteworthy. Adaptive sampling is iterative and accumulates cost over successive iterations. Nevertheless, this does not prevent it from being more efficient, since compared to classical sampling it achieves substantially lower computational times. This may be due to the fact that in each iteration only a small number of samples (15\% of the previous iteration) are added, which can lead to greater efficiency compared to distance-based sampling strategies.

On the other hand, classical methods rely on a single initial exploration of the design space, which results in much higher computational costs. This problem becomes especially challenging in high-dimensional spaces, as the method aims to maximize distances between samples, a task that is computationally expensive. In other words, when using distance-based strategies with large sample sizes, the computational cost of classical sampling grows rapidly, and the method loses efficiency.

When comparing the emission values of $CO_2$ and energy consumption for each sampling method and their corresponding sample sizes and models, we can extrapolate all the results and conclusions obtained for the computational cost in time, as shown in subsection~\ref{sec:CO2-emissions} the corresponding $R^2$ between the temporal cost and emissions and the temporal cost and energy consumption is practically $1$ in both cases. Therefore, carbon emissions are very similar for the $3$ models and time instance experiments for the case of the adaptive sampling, with the values showing negligible variation as the number of samples increases, getting all values within $3.469\cdotp10^{-4} kg$ and $3.972\cdotp10^{-4} kg$, and similar happens for the energy consumption, getting values within $8.501\cdot10^{-3} kWh$ and $9.733\cdotp10^{-3} kWh$. In the case of the classical method, the values are also similar across all models and time-instance experiments; however, both carbon emissions and energy consumption increase considerably as the number of samples grows, reaching values between $3.493\cdotp10^{-4}kg$ and $2.387\cdotp10^{-2}kg$ for the $CO_2$ emissions and values between $8.561\cdotp10^{-3} kWh$ and $5.849\cdotp10^{-1} kWh$ for the energy consumption.

The results clearly show this effect: classical sampling struggles to identify spatially scattered sets of distant samples, which has negative consequences for computational efficiency. By contrast, adaptive sampling, through its iterative construction of the design space and strategic selection of new samples, keeps computational costs under control.

\subsubsection{Training temporal cost}

When training surrogate models with classical and adaptive sampling, it should be noted that in the former case, the model is trained only once, whereas in the latter, training is repeated for each newly added batch of samples.

Therefore, for adaptive sampling, the actual time required to train the $k$-th surrogate model must consider the cumulative computational cost of all models trained from the initial sample set up to the $k$-th iteration.

\begin{figure}[htbp]
    \centering

    \begin{subfigure}[b]{\textwidth}
        \centering
        \includegraphics[width=0.8\textwidth]{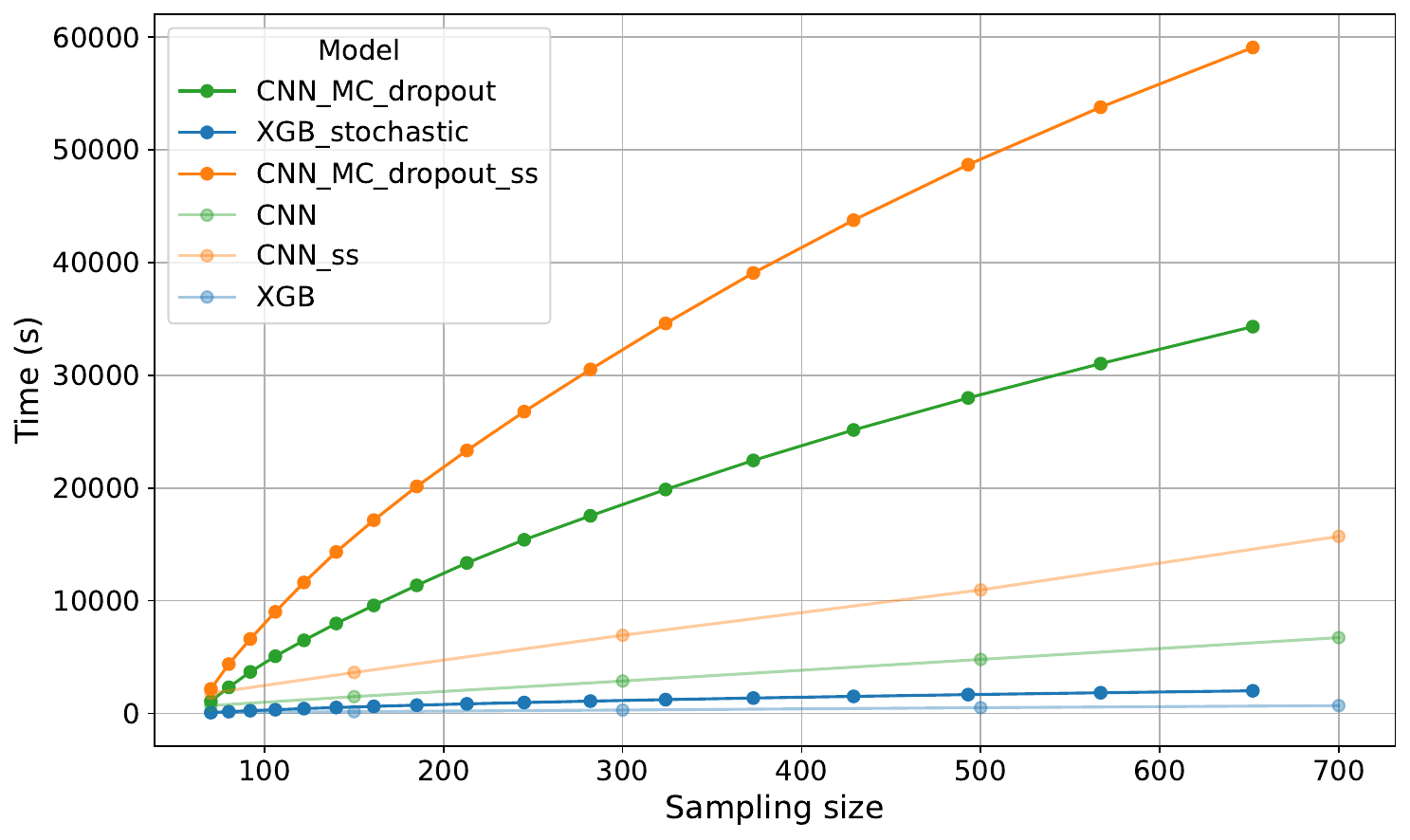}
        \caption{Training cost of surrogate models for training instance ID1.}
        \label{fig:entrenamendua-sub1}
    \end{subfigure}
    \hfill
    \begin{subfigure}[b]{\textwidth}
        \centering       
        \includegraphics[width=0.8\textwidth]{figures/t_training_cost_ID1.pdf}
        \caption{Training cost of surrogate models for training instance ID2.}
        \label{fig:entrenamendua-sub2}
    \end{subfigure}
    \hfill
    \begin{subfigure}[b]{\textwidth}
        \centering
        \includegraphics[width=0.8\textwidth]{figures/t_training_cost_ID1.pdf}
        \caption{Training cost of surrogate models for training instance ID3.}
        \label{fig:entrenamendua-sub3}
    \end{subfigure}

    \caption{Computational cost of training surrogate models for each group of training instances, for both sampling methods.}
    \label{fig:entrenamendu-kostea}
\end{figure}

The computational costs for all models, across different sample sizes and both sampling methods, are shown in Figure \ref{fig:entrenamendu-kostea}.

Comparing the results across the three training instance experiments, it is evident that the trends among models are consistent. It is also important to highlight that the computational cost depends on the amount of data and the number of parameters to optimize. Although the trends remain similar across the three instance sets, an increase in the number of training instances directly translates into higher computational cost. This effect is particularly pronounced in neural networks. For example, training a convolutional neural network with classical sampling for $ID1$ requires approximately $7000s$, whereas for the same model, the cost rises to between $10000$ and $11000s$ for $ID2$ and $ID3$.

\begin{figure}[htbp]
    \centering

    \begin{subfigure}[b]{\textwidth}
        \centering
        \includegraphics[width=0.87\textwidth]{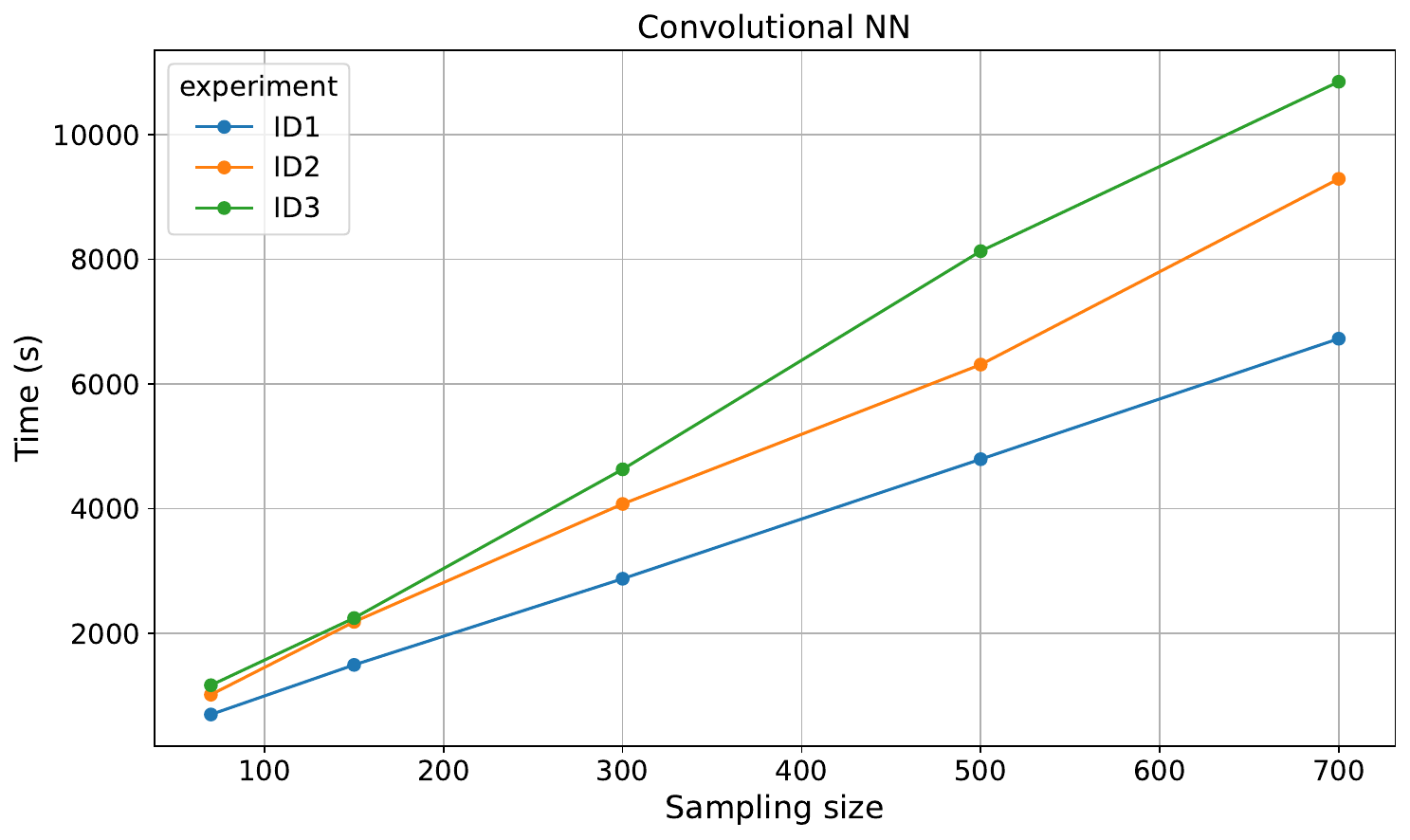}
        \caption{Computational cost of convolutional neural network training.}
        \label{fig:CNN-sub1}
    \end{subfigure}
    \hfill
    \begin{subfigure}[b]{\textwidth}
        \centering       
        \includegraphics[width=0.87\textwidth]{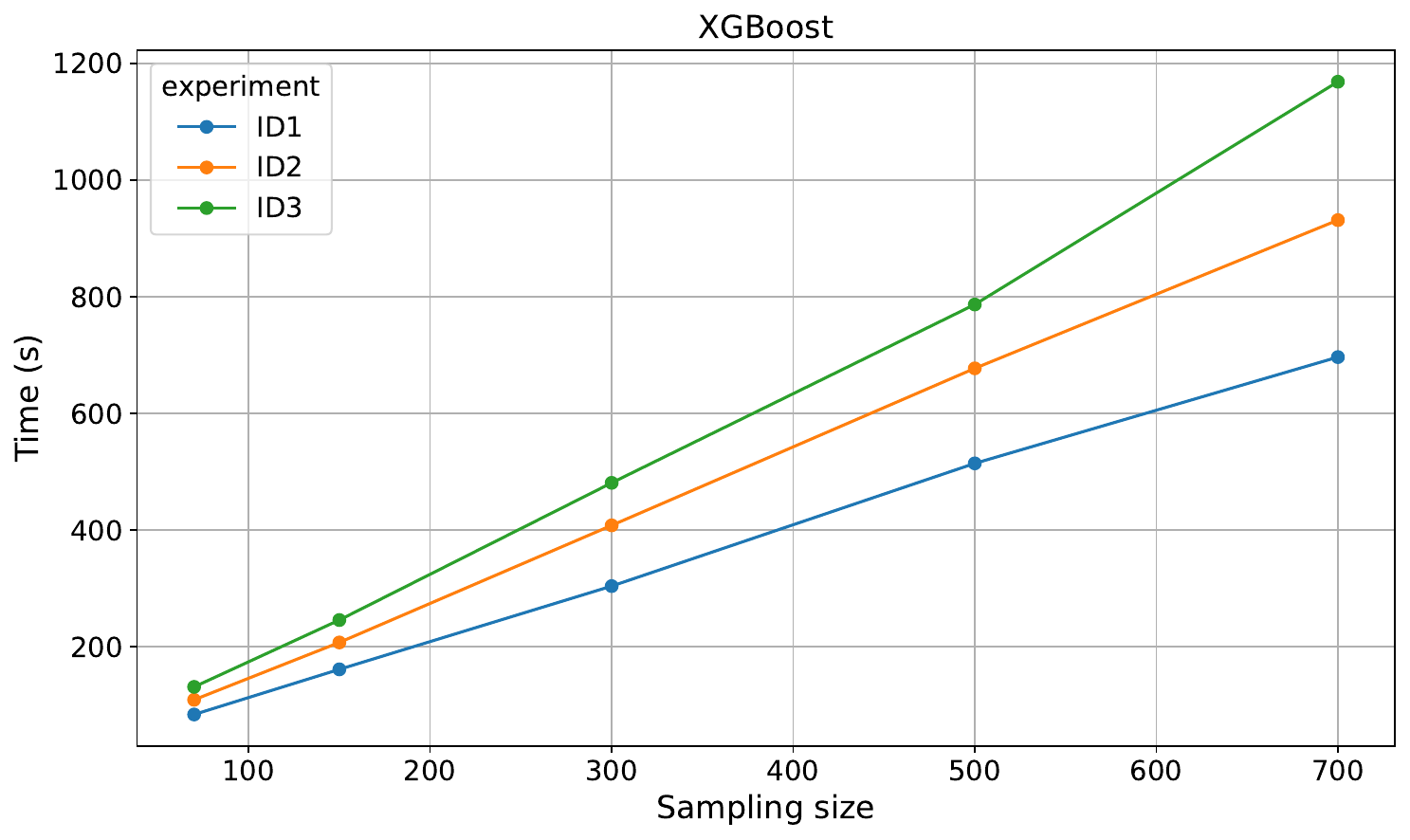}
        \caption{Computational cost of XGBoost training.}
        \label{fig:XGB-sub2}
    \end{subfigure}

    \caption{Training computational cost of surrogate models across different training instance sets.}
    \label{fig:entrenamendu-kostea-ereduka}
\end{figure}

The variation of computational cost across the three training instance sets is illustrated in Figure \ref{fig:entrenamendu-kostea-ereduka} for convolutional neural networks and XGBoost. Similar trends are observed for the remaining models.

Again, referring to Figure \ref{fig:entrenamendu-kostea}, it is clear that, in general, training with adaptive sampling incurs higher costs compared to classical sampling. As expected, cumulative training costs increase iteratively with each added batch of samples, whereas classical sampling trains only once.

However, in convolutional neural networks, the relationship between sample size and computational time is approximately linear for classical sampling, but not for adaptive sampling. Specifically, adaptive sampling exhibits a concave curve, with the derivative decreasing as more samples are added. This effect is attributed to incremental training: although cross-validation is required to estimate uncertainty, the model is ultimately retrained with all data incrementally, which reduces the effective training time for each batch of new samples.

In the case of the environmental cost associated with the training phase, as with the sampling cost, the results and conclusions are also directly extrapolable, because of the $R^2$ values obtained in subsection~\ref{sec:CO2-emissions}. More precisely, in these cases, both $CO_2$ emissions and energy consumption rise as the number of instances in the temporal experiments increases. In addition, it can be observed that the results of the adaptive method are higher and similar across all models, whereas in the classical method the increase is barely noticeable in comparison. All this leads to carbon dioxide emissions values between $1.888\cdotp10^{-3}kg$ and $8.901\cdotp10^{-2}$ in the case of adaptive sampling, and an energy consumption between $4.626\cdotp10^{-2}kWh$ and $2.181kWh$ for the same sampling method. In the case of the classical sampling, the $CO_2$ emissions are between $1.202\cdotp10^{-3}kg$ and $1.866\cdotp10^{-2}kg$, and energy consumptions between $2.945\cdotp10^{-2}kWh$ and $4.573\cdotp10^{-1}kWh$. 

Finally, it is noteworthy that XGBoost exhibits particularly low computational costs compared to neural networks. Although adaptive sampling increases the training cost, it still improves performance relative to classical sampling in neural networks, achieving better accuracy for the same or lower computational expense.

\subsection{Models performance}\label{sec:performance}

In this section, the performance metrics obtained for each model are compared. Specifically, the coefficient of determination ($R^2$) was used to evaluate the fitting accuracy of the surrogate models.

\begin{figure}[htbp]
    \centering

    \begin{subfigure}[b]{\textwidth}
        \centering
        \includegraphics[width=0.8\textwidth]{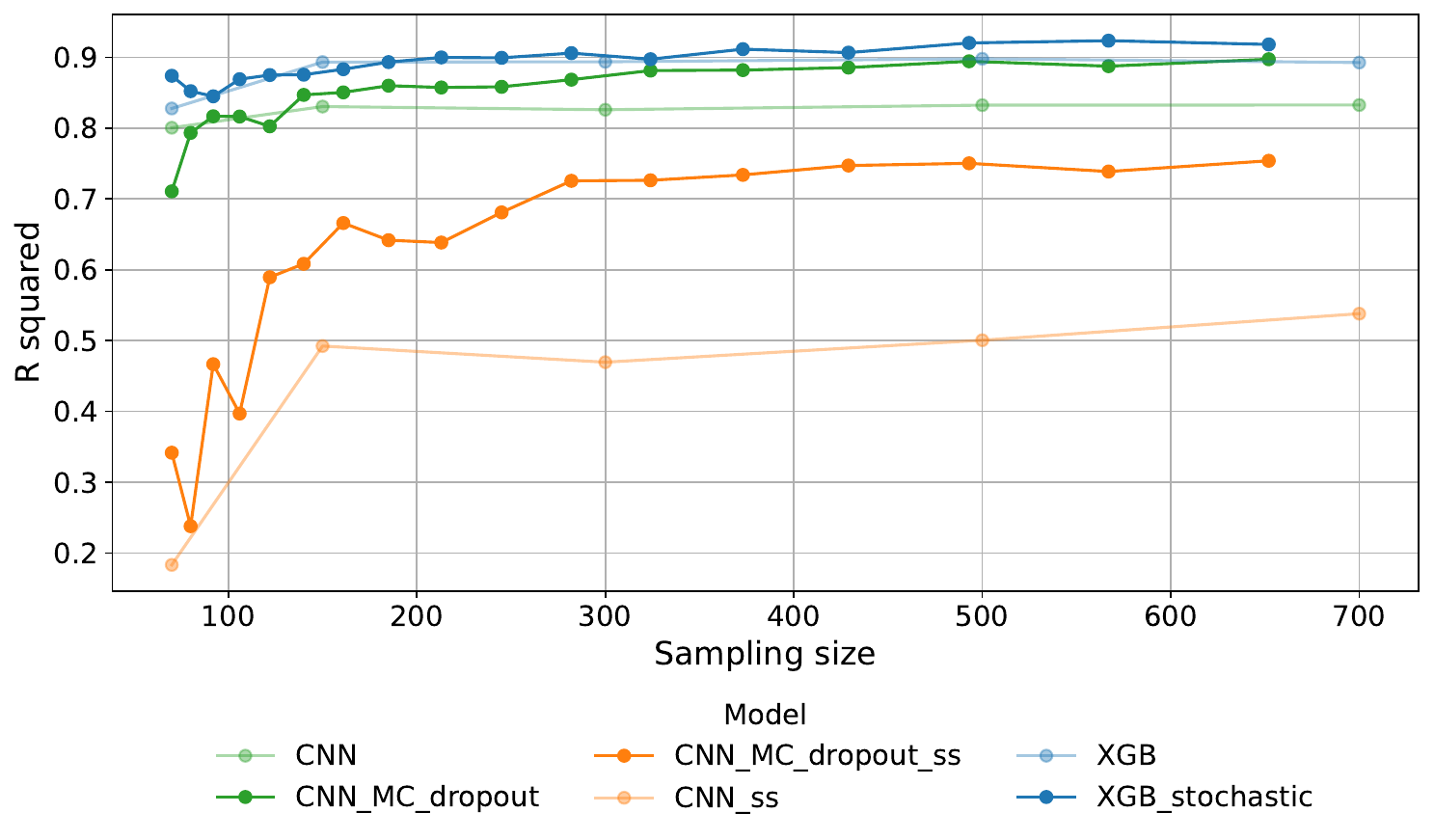}
        \caption{Model performance metrics for training instance set ID1.}
        \label{fig:metrikak-sub1}
    \end{subfigure}
    \hfill
    \begin{subfigure}[b]{\textwidth}
        \centering       
        \includegraphics[width=0.8\textwidth]{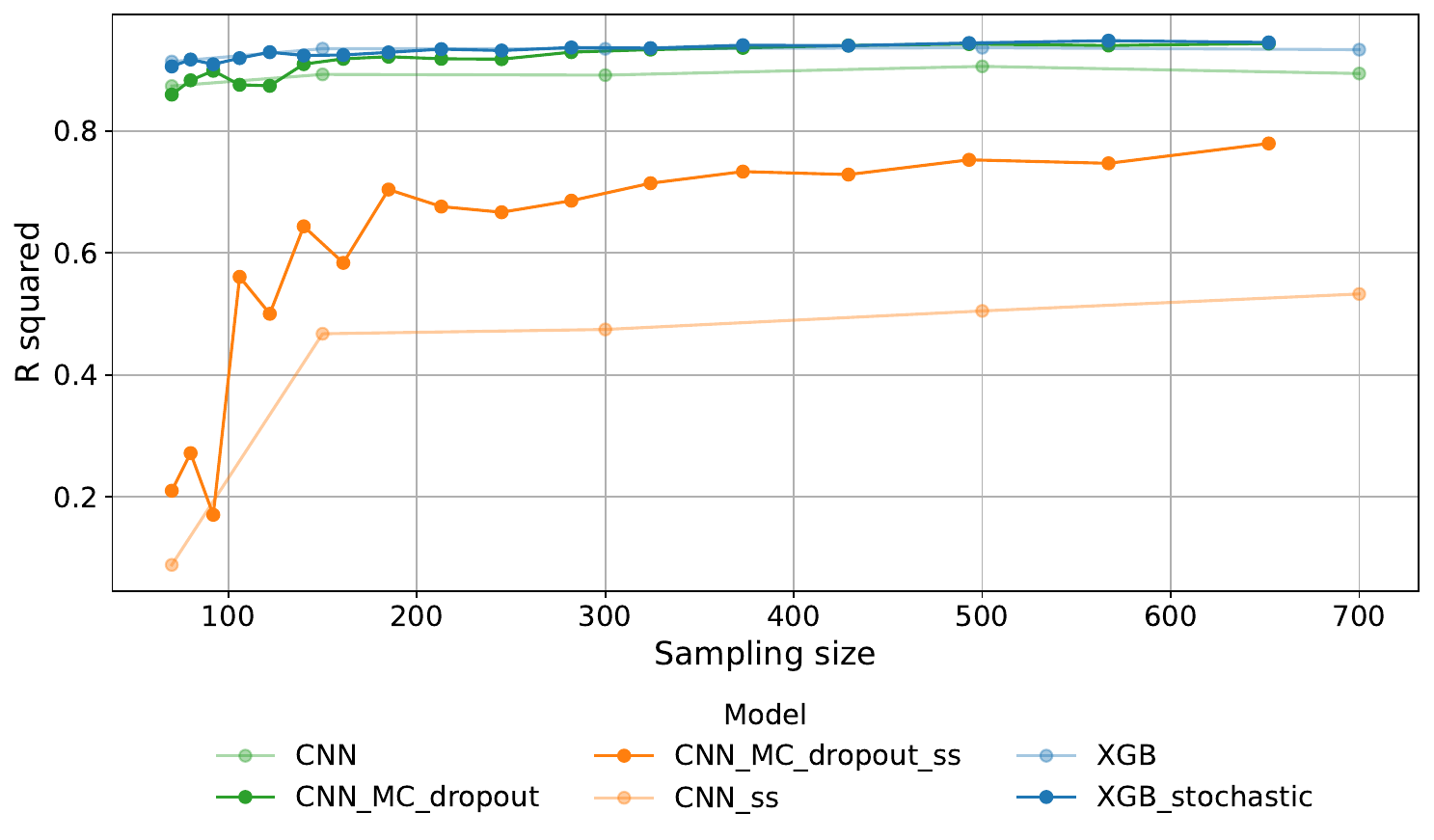}
        \caption{Model performance metrics for training instance set ID2.}
        \label{fig:metrikak-sub2}
    \end{subfigure}
    \hfill
    \begin{subfigure}[b]{\textwidth}
        \centering
        \includegraphics[width=0.8\textwidth]{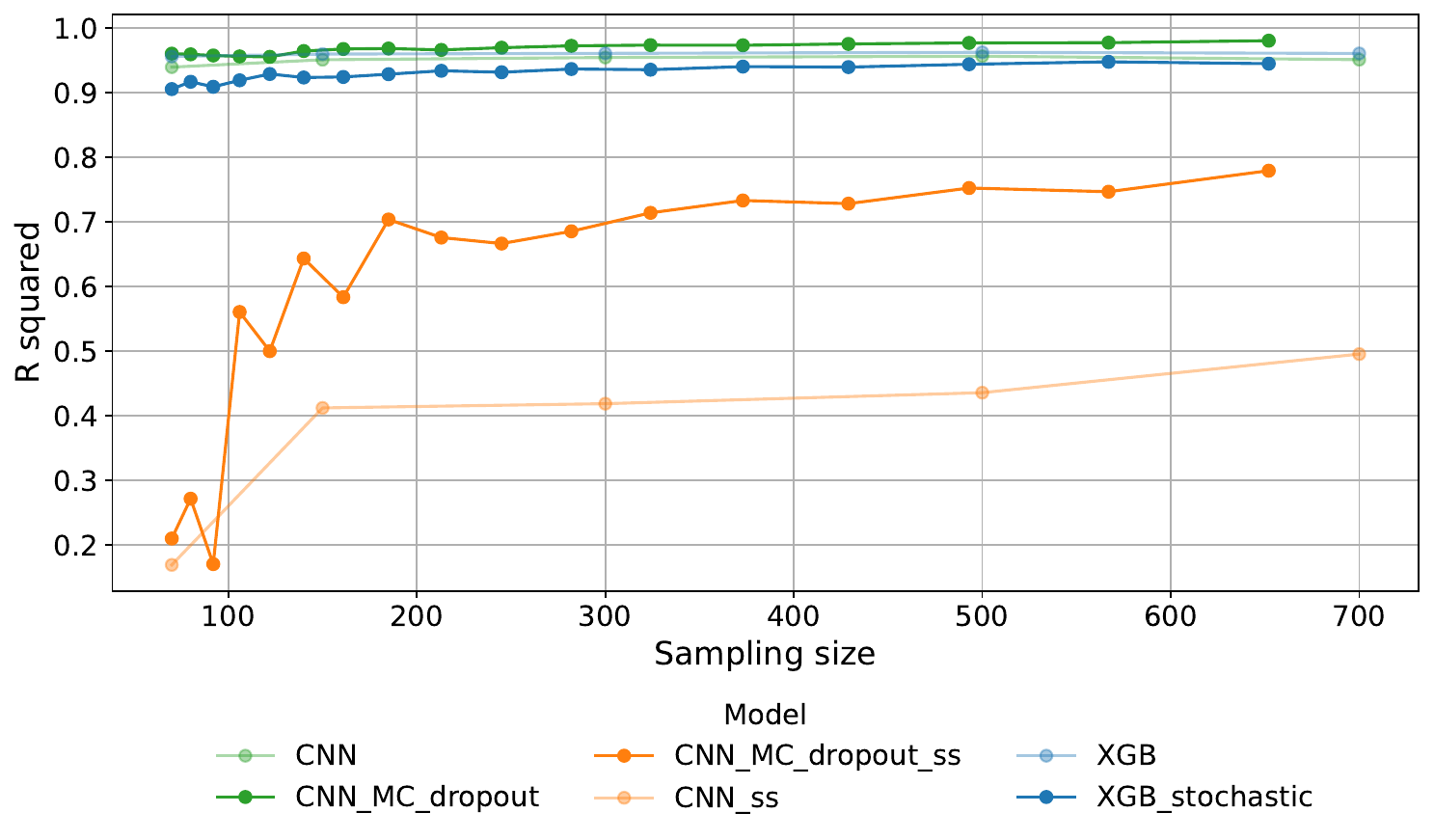}
        \caption{Model performance metrics for training instance set ID3.}
        \label{fig:metrikak-sub3}
    \end{subfigure}

    \caption{Fitting metrics obtained for each model across training instance sets and both sampling methods.}
    \label{fig:metrikak-ereduak}
\end{figure}

The same plots used for comparing computational costs across models, training instance sets, and sampling methods are employed here. These results are shown in Figure \ref{fig:metrikak-ereduak}.

As observed in Figure \ref{fig:metrikak-ereduak}, self-attentive convolutional neural networks do not achieve high performance, with $R^2$ values remaining below $0.80$ regardless of the sampling method or training instance set. However, adaptive sampling clearly improves performance compared to classical sampling. Specifically, no classical sampling experiment achieves $R^2 > 0.50$, while adaptive sampling yields values between $0.70$ and $0.80$ across the three training instance sets.

For convolutional neural networks, adaptive sampling consistently outperforms the self-attentive variant across all cases. Moreover, for all three training instance sets, adaptive sampling enhances the results obtained with classical sampling. The difference in $R^2$ between the two methods decreases as the last training instance increases. Beyond reducing the difference, both models achieve higher performance metrics as the number of training instances grows. Notably, the best performance is obtained for $ID3$, while results in other experiments remain very close to the highest values.

In the case of XGBoost, excellent results are obtained across all three training instance sets. Specifically, the best outcomes occur for $ID1$ and $ID2$, and for $ID3$, XGBoost performance closely approaches that of convolutional neural networks. Regarding the sampling methods, adaptive sampling improves results for $ID1$ relative to classical sampling. For $ID2$, both methods yield very similar performance, and for $ID3$, classical sampling slightly outperforms adaptive sampling.

Finally, it is important to note that both convolutional neural networks and XGBoost achieve satisfactory initial performance, with $R^2 > 0.7$ in all cases. As the sample size increases, fitting metrics improve gradually, approaching nearly perfect regression in some instances. However, the rate of improvement is relatively slow. 

\subsection{Discussion}\label{sec:discussion}

Following the hypotheses and objectives outlined earlier, the following key points should be considered when selecting an appropriate surrogate model:

\begin{itemize}
    \item Minimize the number of training instances whenever possible.
    \item Ensure that the last training instance used is sufficiently far from the final prediction instance.
    \item Achieve high fitting metrics with the smallest possible number of samples.
    \item Select a model that balances training cost and fitting performance.
\end{itemize}

As observed in previous results, increasing the number of training instances leads to higher training costs, particularly for neural networks.

Regarding the last training instance used, this choice carries significant weight. Instances closer to the prediction scenario tend to yield better fitting metrics. However, when deploying the surrogate model, greater dependency on the physical model may arise, potentially reducing the overall efficiency of the methodology. For example, in the case of $ID1$, an efficiently trained surrogate can reduce up to 50\% of the original model’s computational time for predictions. For $ID2$, this saving reaches up to 37.5\%, and for $ID3$, only up to 12.5\%.

The number of samples used for training is critical in scenarios where the physical model is computationally demanding. For instance, if one model requires 300 samples to reach $R^2 = 0.90$ and another only 150, the sampling cost can be reduced by 50\%. When simulations require hours or days, achieving high accuracy with fewer samples is particularly important.

Finally, the balance between training cost and model accuracy must be maintained. Even if all previous conditions are satisfied, the training process must remain efficient; otherwise, the surrogate model will not offer meaningful computational savings relative to the original model.

\begin{figure}[htbp]
    \centering

    \begin{subfigure}[b]{\textwidth}
        \centering
        \includegraphics[width=0.84\textwidth]{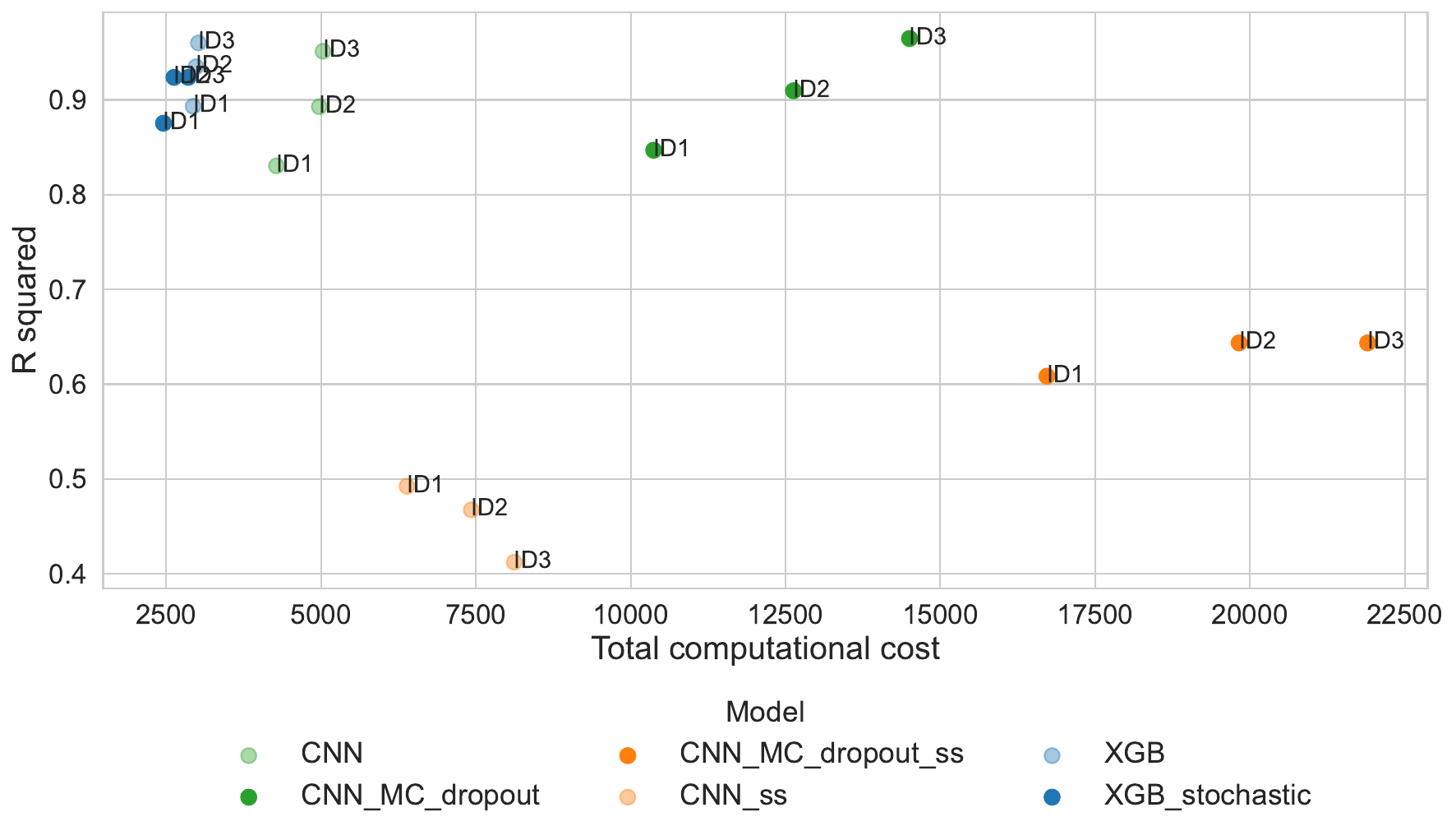}
        \caption{Total computational cost and $R^2$ for 150 samples}
        \label{fig:konparazioa-sub1}
    \end{subfigure}
    \hfill
    \begin{subfigure}[b]{\textwidth}
        \centering       
        \includegraphics[width=0.84\textwidth]{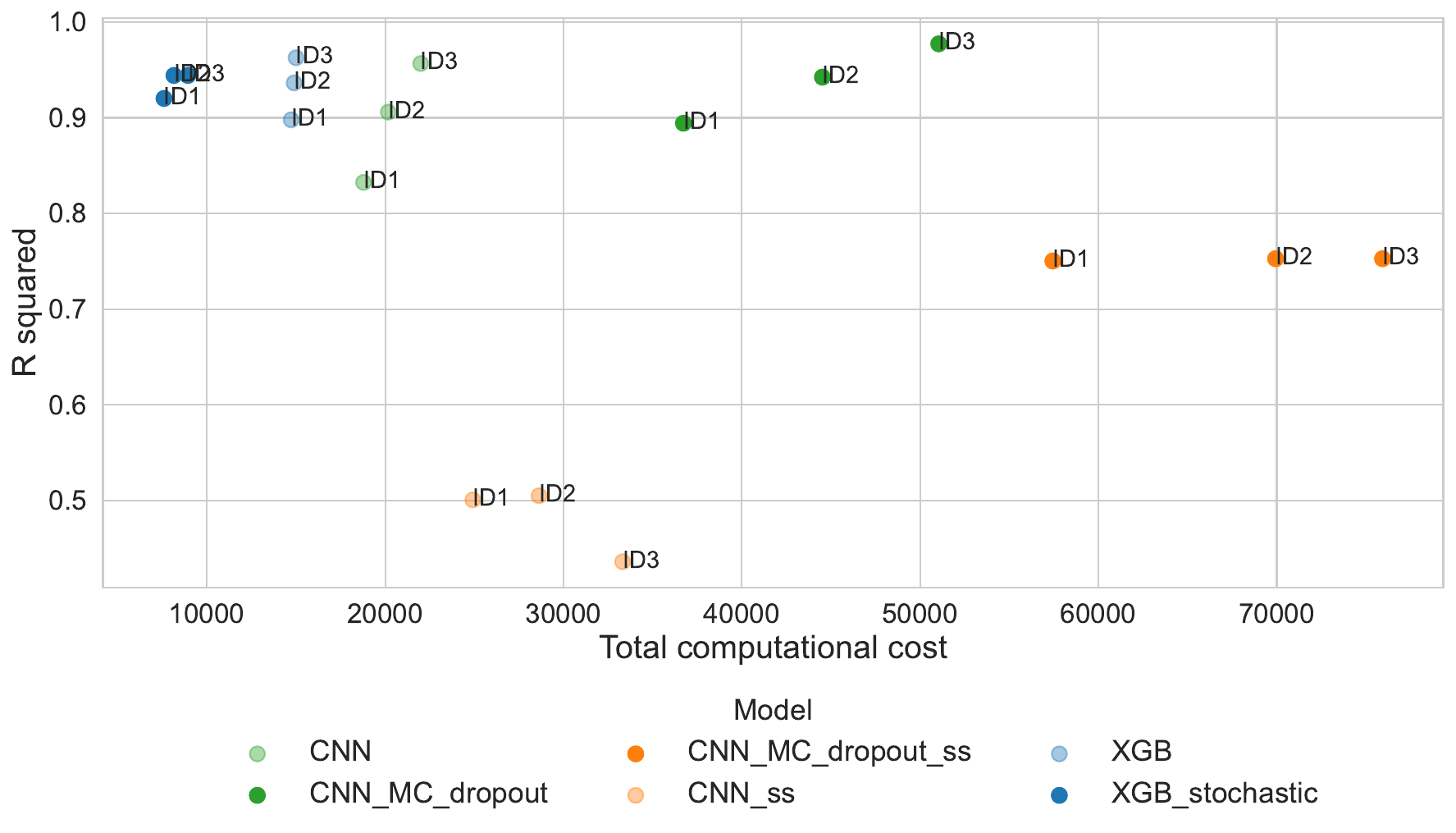}
        \caption{Total computational cost and $R^2$ for 500 samples}
        \label{fig:konparazioa-sub2}
    \end{subfigure}
    \hfill
    \begin{subfigure}[b]{\textwidth}
        \centering
        \includegraphics[width=0.84\textwidth]{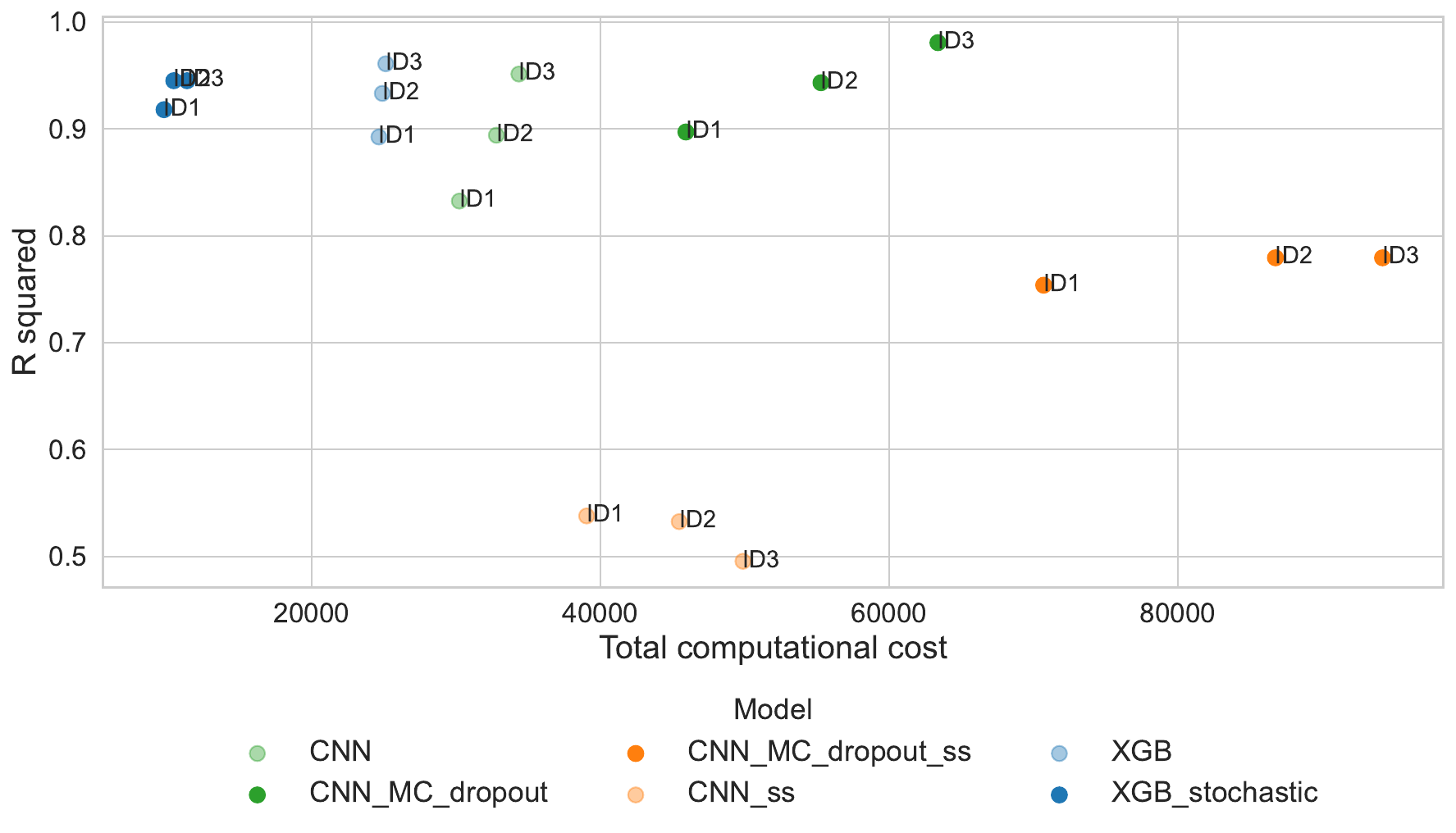}
        \caption{Total computational cost and $R^2$ for 700 samples}
        \label{fig:konparazioa-sub3}
    \end{subfigure}

    \caption{Trade-off between total computational cost and $R^2$ across training instance sets and models for different sample sizes.}
    \label{fig:adaptive-konparazioa}
\end{figure}

Figure \ref{fig:adaptive-konparazioa} illustrates the balance between the number of training instances, the last training instance, and the trade-off between fitting accuracy and computational cost for 150, 500, and 700 samples. For adaptive sampling, results closest to these sample sizes are shown. Total computational cost includes sampling, simulation, and training; simulation cost is effectively constant for the same sample size across models. Thus, the cost reflects the trade-off between sampling and training.

Models that achieve a good balance between accuracy and computational cost include XGBoost with either classical or adaptive sampling, and convolutional neural networks with classical sampling. For convolutional networks with adaptive sampling, computational cost increases significantly with sample size, but better results are obtained with fewer samples compared to classical sampling, maintaining the trade-off.

Regarding training instances, $ID3$ experiments for XGBoost are unnecessary, as results are very similar to $ID2$, while prediction savings are much larger. Between $ID1$ and $ID2$, the choice should depend on simulation cost. For high-cost simulations, the accuracy offered by $ID1$ may be sufficient considering prediction savings; if higher fitting is prioritized, $ID2$ is preferable.

For convolutional networks, adaptive sampling increases computational cost, but small sample sizes ($ID1$ and $ID2$) achieve good results, especially when considering the prediction cost of $ID3$. Specifically, with 150 samples and adaptive sampling for $ID2$, computational cost and fitting metrics are much better than those obtained with 300 samples using classical sampling, as shown in Figures \ref{fig:konparazioa-sub1} and \ref{fig:konparazioa-sub2}.

Self-attentive convolutional networks are generally not suitable as surrogate models, as the trade-off between computational cost and fitting accuracy is poor in all cases.

A main objective is to train a surrogate model that achieves high fitting metrics with minimal sample size. Following literature findings, this work hypothesizes that adaptive sampling allows training a more accurate model with fewer samples.

To test this hypothesis, a comparison was performed between the number of samples and computational cost required to achieve the maximum $R^2$ with classical sampling versus adaptive sampling.

\begin{figure}[htbp]
    \centering

    \begin{subfigure}[b]{\textwidth}
        \centering
        \includegraphics[width=0.95\textwidth]{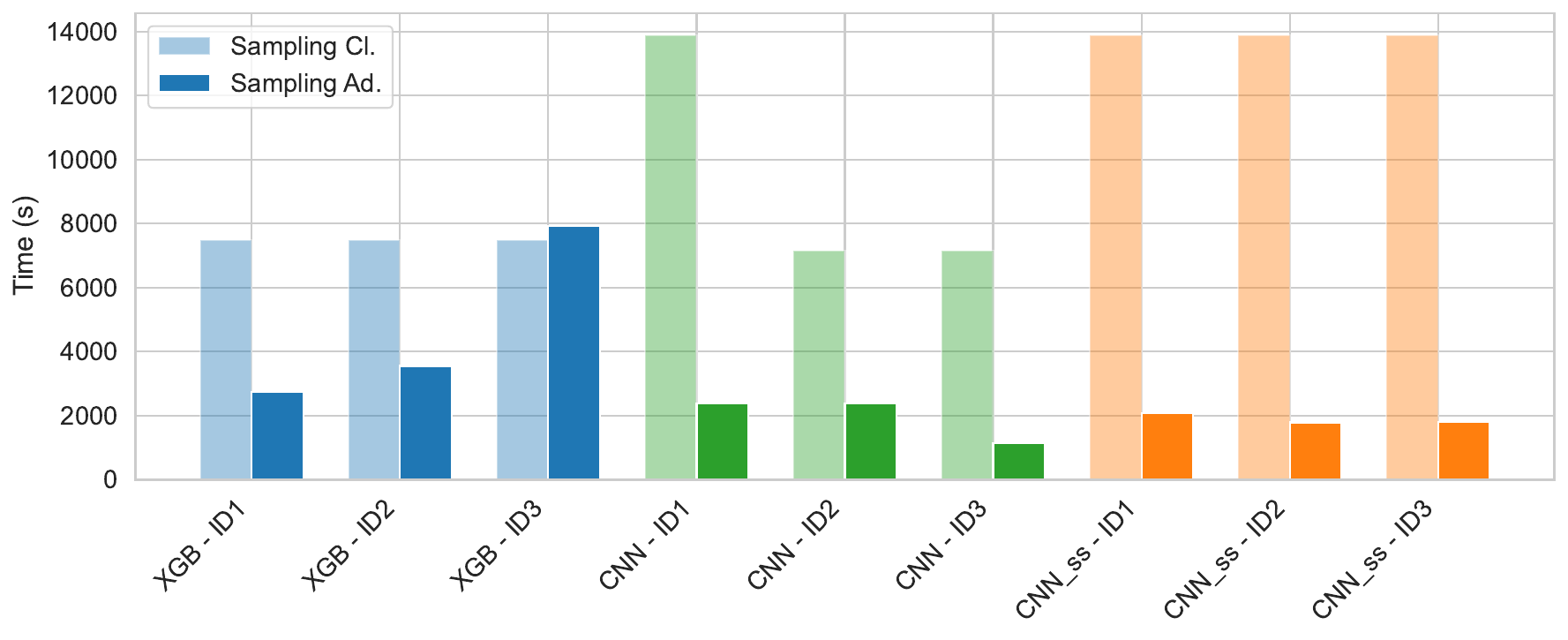}
        \caption{Sampling cost to reach maximum classical $R^2$}
        \label{fig:same-metric-konparazioa-sub1}
    \end{subfigure}
    \hfill
    \begin{subfigure}[b]{\textwidth}
        \centering       
        \includegraphics[width=0.95\textwidth]{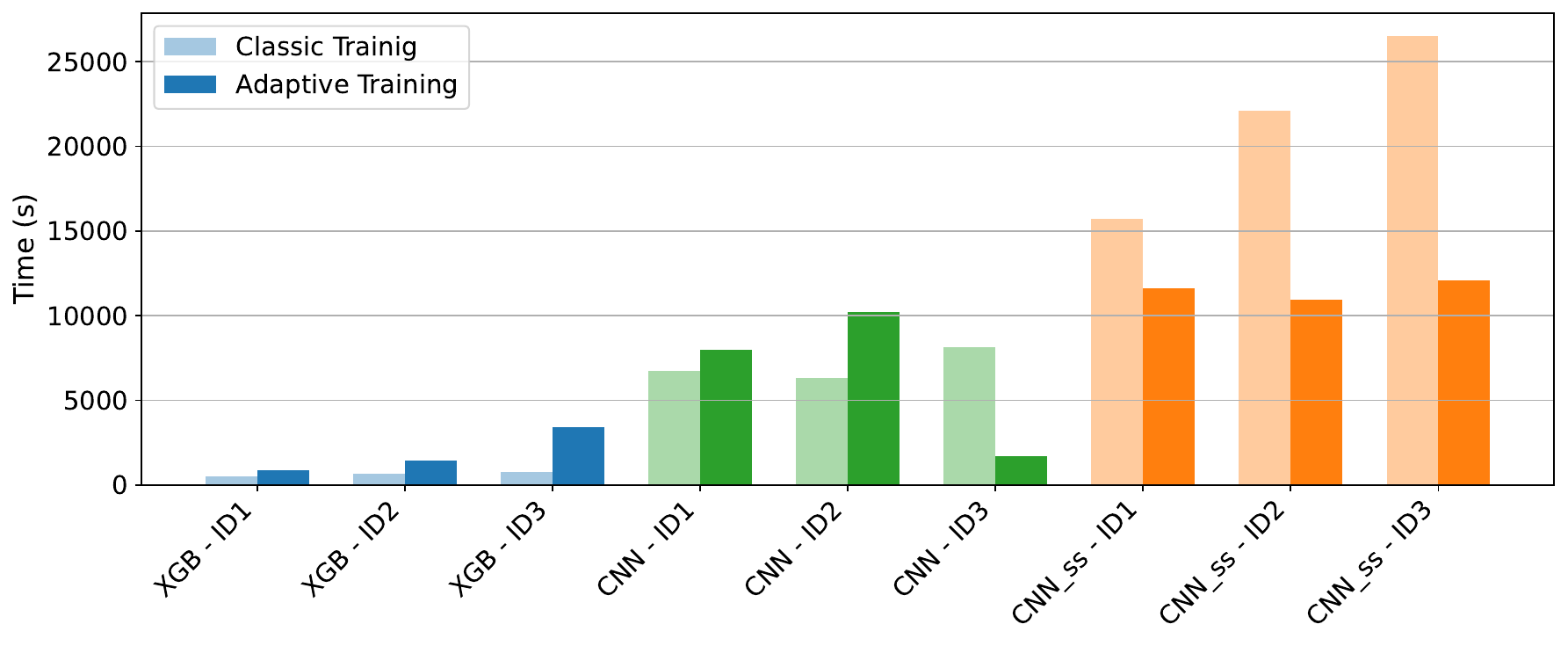}
        \caption{Training cost to reach maximum classical $R^2$}
        \label{fig:same-metric-konparazioa-sub2}
    \end{subfigure}
    \hfill
    \begin{subfigure}[b]{\textwidth}
        \centering
        \includegraphics[width=0.95\textwidth]{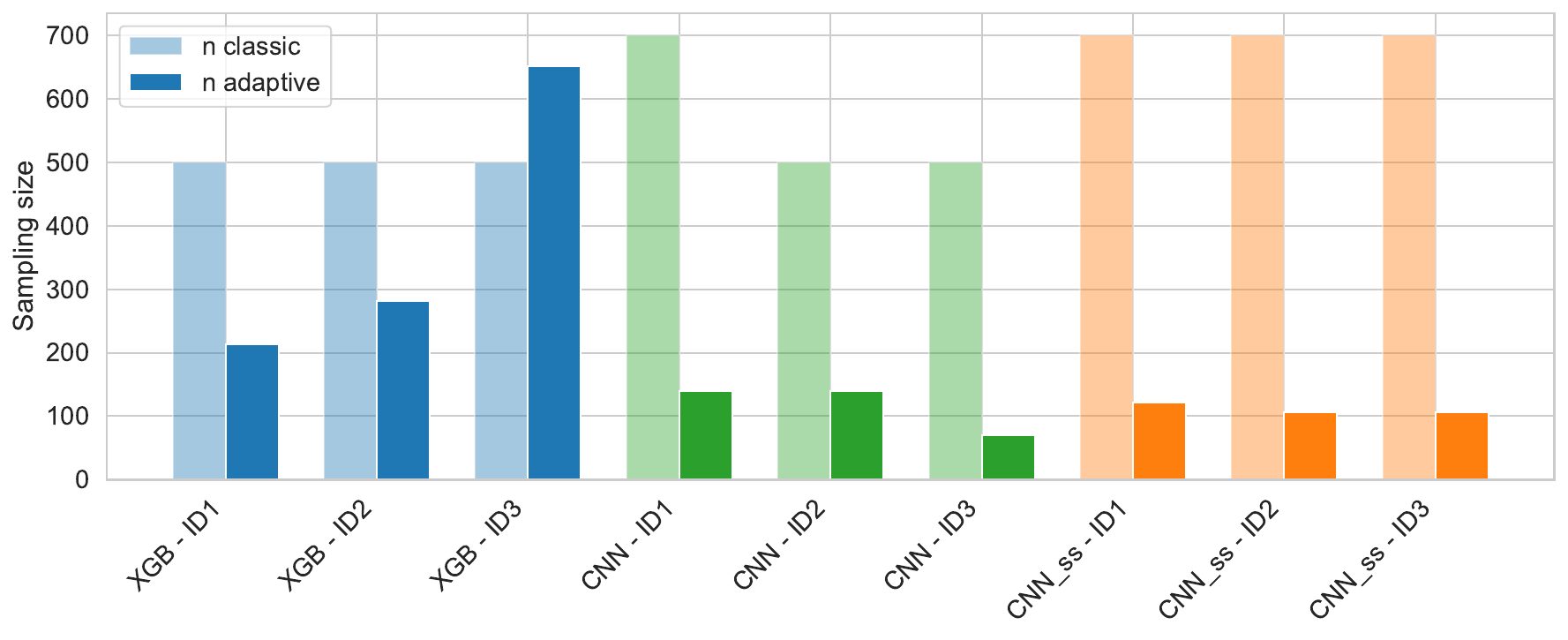}
        \caption{Comparison of sample numbers to reach maximum classical $R^2$}
        \label{fig:same-metric-konparazioa-sub3}
    \end{subfigure}

    \caption{Computational costs in sampling and training, and number of samples, required by each method to reach the maximum $R^2$ achieved with classical sampling.}
    \label{fig:same-metric-konparazioa}
\end{figure}

The results are summarized in Table \ref{tab:kosteak} and visually in Figure \ref{fig:same-metric-konparazioa}.

\begin{table}[htbp]
\centering
\resizebox{\textwidth}{!}{
\begin{tabular}{|l|l|c|c|c|c|c|c|c|}
\hline
\textbf{Model} & \textbf{ID} & \textbf{Max $R^2$ (classical)} & \textbf{n (classical)} & \textbf{Train t (classical)} & \textbf{Sample t (classical)} & \textbf{n (adaptive)} & \textbf{Train t (adaptive)} & \textbf{Sample t (adaptive)} \\
\hline
XGB     & ID1 & 0.899 & 500 & 514.18  & 7479.67 & 213 & 850.77  & 2750.05 \\
XGB     & ID2 & 0.936 & 500 & 677.12  & 7479.67 & 282 & 1426.40   & 3553.86 \\
XGB     & ID3 & 0.963 & 500 & 786.60  & 7479.67 & 652 & 3434.85  & 7919.31 \\
CNN     & ID1 & 0.833 & 700 & 6726.77  & 13877.10 & 140 & 7981.12  & 2390.51 \\
CNN     & ID2 & 0.906 & 500 & 6311.46  & 7145.37 & 140 & 10237.30  & 2392.34 \\
CNN     & ID3 & 0.957 & 500 & 8131.22  & 7145.37 & 70  & 1727.71  & 1139.70 \\
CNN\_ag & ID1 & 0.538 & 700 & 15707    & 13879.80 & 122 & 11618.60  & 2075.11 \\
CNN\_ag & ID2 & 0.533 & 700 & 22119.40  & 13879.80 & 106 & 10954.80  & 1788.75 \\
CNN\_ag & ID3 & 0.496 & 700 & 26538.50  & 13879.80 & 106 & 12092.20  & 1791.29 \\
\hline
\end{tabular}}
\caption{Computational cost and number of samples required by each method to reach the maximum classical $R^2$.}
\label{tab:kosteak}
\end{table}

As shown in Table \ref{tab:kosteak} and Figure \ref{fig:same-metric-konparazioa}, except for $ID3$ experiments with XGBoost, adaptive sampling requires fewer samples to reach the maximum classical $R^2$, substantially reducing simulation time. For example, in XGBoost with $ID1$, adaptive sampling achieves a 57.4\% saving in simulation cost. On average, across all experiments, 65.7\% of simulation time could be saved with adaptive sampling.

As previously noted, adaptive sampling is computationally more efficient than classical sampling when considering the time to select samples. Combined with achieving the same accuracy with fewer samples (Figure \ref{fig:same-metric-konparazioa-sub2}), adaptive sampling provides significantly better computational efficiency in most cases.

Finally, although the cumulative computational cost of models trained with adaptive sampling is higher than that of classical sampling, reaching the same fitting metrics with fewer samples reduces the overall cost, as shown in Figure \ref{fig:same-metric-konparazioa-sub1}. Except for self-attentive convolutional networks, training costs are higher but more balanced across the remaining models.

\section{Conclusions}\label{sec:conclusions}

The main goal of this work has been to identify an appropriate surrogate modelling strategy for the spatio-temporal nature of dendritic solidification simulations, with the aim of reducing the computational cost of the original phase field model. To this end, the study has examined how different modelling choices, the selection of temporal training instances, the sampling methodology, and the model, have influenced the number of required samples, the computational cost and the final predictive performance of the surrogate. Since each sample corresponds to an expensive phase field simulation, minimizing the number of samples has been essential for reducing total computational effort. The evaluation has been conducted through quantitative metrics capturing surrogate accuracy ($R^2$), computational time, training and sampling costs, and their associated energy consumption and $CO_2$ emissions. Training time exhibited an almost perfectly linear correlation with environmental indicators, thus, it has served as a consistent basis for assessing the trade-off between sustainability, computational efficiency, and surrogate-model performance.

The comparison between the classical and adaptive sampling methods proposed revealed clear differences. Classical sampling, based on an Optimal Latin Hypercube Sampling conducted by the PSO algorithm, struggles to remain computationally efficient as the number of samples increases, since the optimization is distance-based and its cost grows accordingly, leading to higher sampling costs. For the proposed adaptive sampling method, the cost associated with identifying new samples is lighter than in the classical approach, because new points are added iteratively and the search process is designed to minimise its computational impact. In addition, the method progressively refines the design space and incorporates uncertainty—estimated through cross-validation—to guide the selection of new samples. While this strategy increases training time, since the model must be retrained at every iteration, it generally reduces the total number of samples required to reach a given accuracy. As a result, adaptive sampling achieves better sample efficiency and reduces the simulation cost in most experiments.

Regarding the comparison between the convolutional models—which automatically extract spatial features—and the XGB model—which relies on a domain-knowledge-based feature transformation—the training cost of the CNNs is consistently higher, as they must optimise a significantly larger number of parameters. Furthermore, as mentioned previously, the total training cost in the adaptive approach is also higher, since it accumulates the training effort required at each iteration. This cost is further increased by the need to estimate model uncertainty across the entire design space: in the current implementation, this is achieved through cross-validation, which adds an additional computational burden to every adaptive iteration.

When comparing the different sets of temporal instances, it can be observed that—except for the self-supervised models—better predictive metrics are obtained when the final time $t_f$ is closer to the target solidification state. This improvement, however, comes at the expense of a slight increase in computational cost. Considering these results together with the simulation time required when using the surrogate for prediction, we conclude that the best balance between performance and computational efficiency is achieved with the $ID1$ and $ID2$ configurations.

Finally, in terms of the trade-off between training cost and predictive performance, the XGB model achieves the best overall balance, largely due to the efficiency of the domain-knowledge-based feature extraction employed. However, although adaptive sampling reduces the number of required samples for both modelling approaches, the reduction is substantially more pronounced for the CNN. This is relevant because, unlike XGB—which benefits heavily from the handcrafted feature transformation and thus depends on prior knowledge of the problem—the CNN learns spatial representations directly from the data. Consequently, when combined with the adaptive sampling strategy, the CNN can achieve competitive performance with significantly fewer samples. This not only enhances its potential for generalisation in scenarios where domain knowledge is limited, but also enables a notable reduction in the number of phase field simulations required, and therefore in the computational cost associated with generating the training data.

Future work would explore several directions aimed at further improving the modelling and training efficiency of the proposed framework. First, incorporating architectures capable of capturing both spatial and temporal correlations, such as recurrent–convolutional hybrids or attention-based spatio-temporal models, would allow a more precise representation of the underlying space-time dynamics without discarding temporal dependencies between frames. Second, strategies that reduce the reliance on the original phase field model during inference would be investigated, for example through autoregressive surrogate formulations that predict successive states from previous surrogate outputs while controlling the accumulated error. Third, the methodology should be evaluated on more computationally demanding phase field formulations, including higher-resolution 2D and fully 3D cases. Fourth, developing more computationally efficient uncertainty-estimation techniques would further reduce the overhead of the adaptive sampling procedure. Finally, physics-informed neural network (PiNN) approaches could be considered as a means to embed physical constraints directly into the surrogate, potentially improving generalisation and reducing the amount of training data required.






\section*{Data Availability}
Data is generated using the phase-field model described in \cite{biner_programming_2017}.

\section*{Code Availability}	
The code used to generate the results in this study will be made publicly available in the future. In the meantime, the code may be shared upon reasonable request to the corresponding author.

\section*{Acknowledgments}
This work has been supported by an Elkartek project funded by the Basque Government for basic research.

\section*{Author Contributions}
Eider Garate-Perez: Conceptualization, Methodology, Software/Coding, Validation, Formal Analysis, Investigation, Data Curation, Writing – Original Draft, Writing – Review \& Editing, Visualization.  
Kerman López de Calle-Etxabe: Conceptualization, Methodology, Resources, Writing – Review \& Editing, Supervision, Funding Acquisition.  
Oihana Garcia: Software/Coding, Validation, Formal Analysis, Investigation, Writing – Original Draft, Writing – Review \& Editing, Visualization.  
Borja Calvo: Conceptualization, Methodology, Writing – Review \& Editing, Supervision.  
Meritxell Gómez-Omella: Conceptualization, Methodology, Software/Coding, Writing – Review \& Editing.  
Jon Lambarri: Software/Coding, Writing – Review \& Editing.

\section*{Competing Interests}
The authors declare no competing interests.

\section*{Preprint notice}
This manuscript is a preprint and has not yet been peer-reviewed.


\printbibliography

\end{document}